\documentclass{aa}  

\usepackage{graphicx}   
\usepackage{amsmath}    
\usepackage{amssymb}    
\usepackage{catchfile}
\usepackage{pdflscape}
\usepackage{xcolor}
\usepackage{longtable}
\usepackage{ltablex}
\usepackage{afterpage}
\usepackage{placeins}
\usepackage{multicol}
\usepackage{multirow}

\usepackage{array}
\newcolumntype{L}{>{\raggedright\arraybackslash}X}

\usepackage{txfonts}
\defcitealias{2019A&A...628A..66L}{L19b}
\defcitealias{2013ApJS..208....9P}{P13}

\usepackage{hyperref}

\newcommand{\aref}[1]{\hyperref[#1]{Appendix~\ref{#1}}}

\begin{document} 

\title{Young nearby open clusters and their luminosity functions}



\author{M. {\v Z}erjal\inst{1,2}\thanks{E-mail: mzerjal@iac.es}
\and
N. Lodieu\inst{1,2}
\and
A. Pérez-Garrido\inst{3}
\and
J. Olivares\inst{1,2,4}
\and
V. J. S. Béjar\inst{1,2}
\and
E. L. Mart{\'{\i}}n\inst{1,2,5}
}

\institute{Instituto de Astrof{\'{\i}}sica de Canarias, E-38205 La Laguna, Tenerife, Spain
\and
Universidad de La Laguna, Dpto. Astrof{\'{\i}}sica, E-38206 La Laguna, Tenerife, Spain
\and
Dpto. F{\'{\i}}sica Aplicada, Universidad Politécnica de Cartagena, 30202 Cartagena, Murcia, Spain
\and
Departamento de Inteligencia Artificial, Universidad Nacional de Educación a Distancia (UNED), c/Juan del Rosal 16, E-28040, Madrid, Spain
\and
Consejo Superior de Investigaciones Cient{\'{\i}}ficas, E-28006 Madrid, Spain
}


 
\abstract
   {
Open clusters are groups of coeval stars sharing properties such as distance and metallicity, and they are key to understanding stellar evolution.
}
   {
Our main goal is to study the evolution of open clusters with a special focus on the universality of the luminosity function.
}
   {
We applied an upgraded version of the convergent point technique on about 50 open clusters. The selection of cluster members was based purely on the exquisite astrometry of the \textit{Gaia}\,DR3 and Hipparcos catalogues in the five-dimensional or full six-dimensional space.
   }
   {
We present updated lists of bona fide members of $\sim$50 open clusters within 500\,pc and younger than 1\,Gyr, exploiting the full depth of the third \textit{Gaia} data release complemented by Hipparcos at the bright end, excluding regions in the Galactic plane.
Our catalogues also are complemented by optical and infrared photometry from the major large-scale public surveys. All the data will be made available on a dedicated webpage with interactive plots and a direct link to Aladin and Vizier hosted at the Centre de Donn\'ees de Strasbourg. 
We derived luminosity functions for all bound clusters and compared them in three age groups of $\sim$50\,Myr, $\sim$150\,Myr, and $\sim$600\,Myr, discussing similarities and differences to constrain their dynamical evolution. 
    }
   {Luminosity functions of clusters at 50\,Myr are more likely similar to each other and show a greater degree of similarity than older clusters. We explain this observation with the universal luminosity function within the volume of our sample (500\,pc). Luminosity functions of clusters with ages similar to the Pleiades or Hyades are more diverse, perhaps due to internal dynamical evolution, but more work is needed to provide additional evidence.
    }


\maketitle
%

\section{Introduction}
Open clusters are gravitationally bound groups of a few tens of stars to over a thousand stars that formed from the same cloud of dust and gas. The similar distance, age, and metallicity of the cluster members means that a group of clusters at different ages can be viewed as snapshots of stellar evolution across time. This makes them an ideal laboratory to study the universality of the initial luminosity and mass functions \citep{1955ApJ...121..161S,1979ApJS...41..513M,kroupa98,1998ApJ...504..835S,chabrier01,kroupa01b,bastian10}, nuclear timescales, and the evolution of the stellar multiplicity rates. 

Over 1100 open clusters were already discovered in our Galaxy \citep{1995ASSL..203..127M,2002A&A...389..871D,2013A&A...558A..53K} before the advent of the \textit{Gaia} astrometric mission \citep{2016A&A...595A...1G}. The arrival of the \textit{Gaia} data releases \citep{2017A&A...601A..19G,2018A&A...616A...1G,2021A&A...649A...1G,2022arXiv220800211G} have resulted in improved lists of cluster members as well as a discovery of new clusters (e.g.\ \citealp{2018A&A...618A..93C,2019A&A...627A..35C,2019AJ....158..122K,2019JKAS...52..145S,2021ApJ...923..129J}), stellar streams \citep{2020MNRAS.492.1370B}, tails, and coronae around clusters \citep{2021A&A...645A..84M,2022ApJ...939...94M}.

Precise parallaxes and proper motions from \textit{Gaia} have also improved the search for members in the kinematic parameter space. Different clustering techniques, in mostly five-dimensional space (i.e., missing radial velocities), have been used to evaluate membership in open clusters. Each of these techniques has their own biases and data quality cuts.
For example, hierarchical density-based spatial clustering of applications with noise (HDBSCAN; \citealp{2017JOSS....2..205M}) helped reveal a number of new clusters, associations, and comoving groups within 1\,kpc \citep{2019AJ....158..122K}. \citet{2022A&A...659A..59T} applied the same method to a search for cluster members up to 50\,pc around their centres in \textit{Gaia}\,EDR3 data and detected vast coronae around almost all the clusters, as well as tidal tails in 71 open clusters. \citet{2018A&A...618A..59C,2019A&A...627A..35C,2020A&A...635A..45C} used DBSCAN in \textit{Gaia}\,DR2 and \citet{2022ApJS..262....7H} used it in \textit{Gaia}\,EDR3 to discover numerous new clusters in the Milky Way. 
Unsupervised photometric membership assignment in stellar clusters (UPMASK; \citealp{2014A&A...561A..57K}) was applied to the proper motions and parallaxes in the TGAS catalogue \citep{2015A&A...574A.115M} by \citet{2018A&A...615A..49C} to characterise 128 known open clusters in the solar neighbourhood.
\citet{2018A&A...618A..93C} used the same method on a compiled list of thousands of known or putative clusters from the literature in the \textit{Gaia}\,DR2 catalogue \citep{2018A&A...616A...1G} to obtain a list of members and cluster parameters for 1229 clusters. 
Recently, \citet{2022A&A...663A.131P} used the same algorithm for the analysis of the 25 most distant ($>$9\,kpc) catalogued open clusters.
In contrast with (H)DBSCAN and UPMASK, 
Extreme Deconvolution Gaussian Mixture Models \citep{2011AnApS...5.1657B} take astrometric uncertainties into account. With this method, \citet{2019A&A...625A.115O} analysed the oldest cluster in the solar neighbourhood, Ruprecht\,147; \citet{2019ApJ...887...19P} discovered a disrupting young open cluster in the halo of the Milky Way; and \citep{2021ApJ...923..129J} provided membership lists for 431 open clusters.
\citet{1998A&A...331...81P} developed a convergent point technique to retrieve the Hyades members in the Hipparcos data \citep{1997A&A...323L..49P}. The membership of this cluster was later improved with the same method using the TGAS catalogue \citep{2018MNRAS.477.3197R} and \textit{Gaia}\,DR2 \citep{2019A&A...623A..35L}.
\citet{2019A&A...628A..66L} improved memberships of the {\ensuremath{\alpha}}\,Per, Pleiades, and Praesepe clusters with the same algorithm.

While stellar mass is the main parameter of the evolution of a star, its determination usually depends on models and their assumptions, except in the case of spectroscopic or eclipsing binaries. Observations have shown that the shape of the initial mass function appears universal among star-forming regions and open clusters, at least formally within the error bars
\citep{2010ARA&A..48..339B}. 
Likewise, the luminosity function constructed from direct observables -- stellar magnitudes that are model-free (unlike stellar masses) -- reflects the cluster's initial mass function, the history of its dynamical evolution, and its age.
For example, when stars fuse their hydrogen fuel and become giants, they increase their luminosity. Similarly, the dynamical evolution causes mass segregation and the gradual evaporation of low-mass members due to the interactions with the external massive objects while moving around the Galaxy.

The main goal of this study is to produce catalogues of members for a volume-limited (distances less than 500\,pc) and age-limited (younger than 1\,Gyr) sample of open clusters outside the Galactic plane, taking advantage of the latest \textit{Gaia} Data Release 3 (DR3; \citealp{2022arXiv220800211G}). We do not make any magnitude cuts to reach the faintest member objects, and we include bright Hipparcos stars that are missing parallaxes and proper motions in \textit{Gaia} in order to reach maximal completeness.
Our aim is to study the evolution of the luminosity functions, dividing our sample of clusters into three age groups. We will make all the catalogues and plots available publicly through an interactive website for the scientific community. 

The paper is organised as follows. Section\,\ref{sec.data} introduces the main source of data and the clusters under scrutiny. Section\,\ref{sec.memberships} describes the procedure to infer membership of any source to a given cluster. We describe the catalogues of members in Section\,\ref{sec.clusters_and_their_members} and provide details about the website with interactive plots in Section\,\ref{sec.website}. We analyse the luminosity functions of the clusters in Section\,\ref{sec.luminosity_functions} and conclude in Section\,\ref{sec.conclusions}.

\section{Data} \label{sec.data}
In this section, we describe the selection procedure to compile a list of clusters, which is the main focus of this study. Then, we continue with the summary of the data to evaluate individual stellar membership for each cluster. This includes stellar positions, distances, proper motions, and radial velocities as well as their mass estimates and supplementary photometric measurements.

\subsection{Selection of clusters}
Our preliminary list of clusters is based on the known clusters from 
\citet{2013A&A...558A..53K}, \citet{2018A&A...618A..93C}, \citet{2021A&A...647A..19T}, and \citet{2021MNRAS.504..356D} as well as the WEBDA website.\footnote{\url{https://webda.physics.muni.cz/navigation.html}} 
To avoid crowded and highly reddened regions of the Milky Way, we only selected clusters that are located outside the Galactic plane (|$b$|\,$>$\,10 deg). We excluded clusters beyond 500 pc due to the increasing parallax uncertainties of their members. We set an upper literature age limit of 1\,Gyr since our objective is to focus on young clusters, and there are not many older open clusters within 500\,pc. For example, in the list of \citet{1995ARA&A..33..381F}, we find one such cluster -- NGC\,752 at $\sim430\,\mathrm{pc}$ with the age of $1.61 \pm 0.03 \pm 0.05 \, \mathrm{Gyr}$ \citep{2023AJ....165....6S}. 
However, we added well-known clusters such as IC\,2602, IC\,2391, and $\alpha$\,Persei. These three clusters are within $b$\,=\,10 deg but have been included because they are nearby and young and have been studied in detail in the literature \citep[e.g.][]{2001A&A...372..862R,2005A&A...436..853L,2010MNRAS.409.1002D,2012MNRAS.426.3403L,2022AJ....163..278N,2019A&A...628A..66L}.
The following clusters/regions 
Sigma\,Orionis \citep{1967PASP...79..433G,2001A&A...377L...9B,2018RNAAS...2R..25C}, 
IC\,348 and NGC\,1333 \citep{2016ApJ...827...52L,2023A&A...671A...1O}, 
Feigelson\,1 (also known as $\epsilon$\,Cha Cluster; \citealp{2021AJ....161...87D}), 
Platais\,2 and 5 \citep{1998AJ....116.2423P}, 
UBC\,19 \citep{2018A&A...618A..59C}, 
Melotte\,227 \citep{1968AJ.....73..556E}, 
Collinder\,65 \citep{2018A&A...615A..12Y},
Collinder\,70 \citep{2008A&A...485..931C}, 
and ASCC\,100 \citep{2005A&A...440..403K} 
were removed because they are not isolated, and many of them are found in very young complex regions (e.g. Taurus, Orion, Upper Scorpius and Rho Ophiuchi) that display an extended structure that is often non-trivial to isolate from the surrounding environment.
This selection left us with 49 clusters, which are presented in a list in Table\,\ref{tab.selected_clusters} with their equatorial and Galactic coordinates, distances, and numbers of initial candidates with and without radial velocities.
All clusters are unique, and a few of them are located relatively close to each other. In rare cases, their members might partially overlap. We explain this issue in Section \ref{sec.catalogs_description}.


\newcommand{\selectedclusters}{ASCC 101 & 288.3179 & 36.3917 & $392.4\,\pm\,3.6$ & 68.0206 & 11.6767 & 93 & 38 & 1 \\
ASCC 41 & 116.7218 & 0.0704 & $293.1\,\pm\,2.5$ & 219.2884 & 12.3447 & 162 & 54 & 1 \\
Alessi 10 & 301.2260 & $-$10.4884 & $433.7\,\pm\,3.6$ & 31.6592 & $-$21.0234 & 112 & 28 & 2 \\
Alessi 13 & 51.4891 & $-$35.8355 & $103.5\,\pm\,0.9$ & 237.5839 & $-$56.1610 & 57 & 27 & 1 \\
Alessi 24 & 260.8358 & $-$62.7086 & $477.8\,\pm\,4.8$ & 328.9885 & $-$14.6079 & 189 & 52 & 3 \\
Alessi 3 & 109.2165 & $-$46.3786 & $275.2\,\pm\,2.6$ & 257.6573 & $-$15.1952 & 88 & 47 & 1 \\
Aveni-Hunter 1 & 354.3071 & 48.3612 & $416.4\,\pm\,5.6$ & 110.4164 & $-$12.6989 & 113 & 29 & 2 \\
Blanco 1 & 0.8825 & $-$29.9150 & $234.3\,\pm\,0.8$ & 15.2846 & $-$79.1200 & 519 & 117 & 1 \\
Collinder 135 & 109.3361 & $-$36.7609 & $293.9\,\pm\,1.8$ & 248.7192 & $-$11.0970 & 212 & 66 & 1 \\
Collinder 350 & 267.0184 & 1.4074 & $366.0\,\pm\,3.3$ & 26.8441 & 14.7183 & 155 & 76 & 1 \\
Gulliver 20 & 273.6913 & 11.1062 & $419.8\,\pm\,5.0$ & 38.8340 & 13.1195 & 80 & 28 & 3 \\
IC 2391 & 130.2486 & $-$53.0266 & $150.5\,\pm\,0.5$ & 270.3989 & $-$6.7794 & 198 & 69 & 3 \\
IC 2602 & 160.7801 & $-$64.3706 & $150.4\,\pm\,0.7$ & 289.6021 & $-$4.8724 & 249 & 88 & 3 \\
IC 4665 & 266.3973 & 5.6721 & $340.9\,\pm\,3.1$ & 30.4945 & 17.2198 & 179 & 52 & 1 \\
Mamajek 1 & 130.6553 & $-$78.3476 & $100.9\,\pm\,1.8$ & 291.8782 & $-$21.2700 & 15 & 10 & 1 \\
Mamajek 2 & 264.3584 & $-$7.9914 & $195.1\,\pm\,2.3$ & 17.0597 & 12.4905 & 62 & 20 & 1 \\
Mamajek 3 & 81.8939 & 6.6251 & $98.5\,\pm\,2.3$ & 197.0021 & $-$15.2903 & 24 & 10 & 1 \\
Mamajek 4 & 276.5463 & $-$50.9501 & $441.2\,\pm\,3.1$ & 343.8369 & $-$17.0053 & 90 & 48 & 1 \\
Melotte 111 & 185.8656 & 26.0623 & $85.5\,\pm\,0.6$ & 221.0263 & 83.6569 & 99 & 62 & 1 \\
Melotte 20 & 51.5358 & 48.9839 & $173.3\,\pm\,0.5$ & 147.3072 & $-$6.4265 & 523 & 173 & 5 \\
Melotte 22 & 56.6517 & 24.0765 & $135.0\,\pm\,0.3$ & 166.5277 & $-$23.6113 & 996 & 358 & 1 \\
Melotte 25 & 66.9682 & 16.5421 & $46.5\,\pm\,0.1$ & 179.6565 & $-$21.7361 & 101 & 73 & 2 \\
NGC 1039 & 40.5334 & 42.7536 & $491.6\,\pm\,1.9$ & 143.6702 & $-$15.6167 & 580 & 219 & 2 \\
NGC 1662 & 72.1526 & 10.8962 & $402.6\,\pm\,2.6$ & 187.7547 & $-$21.1051 & 233 & 97 & 1 \\
NGC 1901 & 79.8301 & $-$68.3214 & $415.2\,\pm\,4.1$ & 278.8643 & $-$33.5280 & 74 & 42 & 1 \\
NGC 2516 & 119.5561 & $-$60.7628 & $407.3\,\pm\,0.6$ & 273.8338 & $-$15.8432 & 2211 & 626 & 1 \\
NGC 2632 & 129.9961 & 19.6375 & $183.4\,\pm\,0.4$ & 205.9118 & 32.3824 & 732 & 324 & 1 \\
Platais 3 & 68.8478 & 71.2787 & $178.0\,\pm\,1.9$ & 138.9824 & 15.7877 & 69 & 32 & 1 \\
Platais 4 & 77.1983 & 22.3997 & $313.5\,\pm\,26.9$ & 180.8842 & $-$10.5194 & 74 & 20 & 1 \\
Stephenson 1 & 283.5958 & 36.8612 & $354.6\,\pm\,2.0$ & 66.8801 & 15.3303 & 321 & 63 & 1 \\
Turner 5 & 142.9664 & $-$36.4349 & $416.7\,\pm\,3.9$ & 264.1163 & 10.9797 & 41 & 22 & 1 \\
UBC 1 & 287.9726 & 56.8465 & $323.2\,\pm\,1.8$ & 87.5620 & 19.7780 & 116 & 43 & 4 \\
UBC 12 & 126.0640 & $-$8.5525 & $437.6\,\pm\,9.4$ & 231.7665 & 16.1980 & 24 & 16 & 4 \\
UBC 159 & 339.0372 & 39.6582 & $489.7\,\pm\,2.7$ & 96.4360 & $-$16.1532 & 78 & 23 & 4 \\
UBC 31 & 61.0929 & 32.8044 & $356.2\,\pm\,1.6$ & 163.2998 & $-$14.5508 & 487 & 110 & 3 \\
UBC 7 & 106.6258 & $-$37.6535 & $273.9\,\pm\,1.7$ & 248.6182 & $-$13.4307 & 167 & 34 & 4 \\
UBC 8 & 84.4349 & 57.1315 & $471.9\,\pm\,2.7$ & 154.8780 & 13.3166 & 144 & 60 & 4 \\
UBC 9 & 276.7157 & 26.4682 & $343.7\,\pm\,3.2$ & 54.5648 & 16.7681 & 106 & 32 & 4 \\
UPK 303 & 39.3939 & 40.6252 & $211.0\,\pm\,2.2$ & 143.8065 & $-$17.9051 & 25 & 7 & 4 \\
UPK 305 & 40.3849 & 38.8866 & $400.1\,\pm\,10.2$ & 145.3155 & $-$19.1525 & 39 & 13 & 3 \\
UPK 350 & 101.7852 & 48.6801 & $426.8\,\pm\,4.4$ & 167.3346 & 19.3564 & 60 & 16 & 3 \\
UPK 385 & 75.5438 & 12.9241 & $313.0\,\pm\,5.7$ & 187.9761 & $-$17.2222 & 57 & 11 & 3 \\
UPK 533 & 123.2414 & $-$52.4858 & $336.6\,\pm\,3.0$ & 267.5120 & $-$9.9585 & 62 & 19 & 3 \\
UPK 540 & 114.8198 & $-$58.6372 & $362.2\,\pm\,3.1$ & 270.7512 & $-$17.0598 & 85 & 19 & 3 \\
UPK 606 & 215.7553 & $-$46.4112 & $168.3\,\pm\,2.1$ & 318.9521 & 13.5922 & 36 & 19 & 3 \\
UPK 612 & 254.1704 & $-$68.3064 & $225.1\,\pm\,1.5$ & 322.5872 & $-$15.4121 & 74 & 25 & 3 \\
UPK 624 & 254.3194 & $-$60.5082 & $311.8\,\pm\,4.0$ & 329.0544 & $-$10.8109 & 39 & 14 & 3 \\
UPK 654 & 277.2507 & $-$34.5111 & $496.0\,\pm\,6.3$ & 359.5211 & $-$10.7585 & 52 & 13 & 3 \\
UPK 88 & 309.8308 & 21.1573 & $285.2\,\pm\,4.1$ & 64.6697 & $-$12.2315 & 32 & 8 & 4 \\}
\begin{table*}
\caption{List of 49 clusters analysed in this work.}
\begin{tabular}{l rr rr r rr r}
\hline
Cluster & RA & DEC & Distance & l & b & N(init) & N(RV) & Ref \\
\hline
 & deg & deg & pc & deg & deg &  &  &  \\
 \hline
 ASCC 101 & 288.3179 & 36.3917 & $392.4\,\pm\,3.6$ & 68.0206 & 11.6767 & 93 & 38 & 1 \\
ASCC 41 & 116.7218 & 0.0704 & $293.1\,\pm\,2.5$ & 219.2884 & 12.3447 & 162 & 54 & 1 \\
Alessi 10 & 301.2260 & $-$10.4884 & $433.7\,\pm\,3.6$ & 31.6592 & $-$21.0234 & 112 & 28 & 2 \\
Alessi 13 & 51.4891 & $-$35.8355 & $103.5\,\pm\,0.9$ & 237.5839 & $-$56.1610 & 57 & 27 & 1 \\
Alessi 24 & 260.8358 & $-$62.7086 & $477.8\,\pm\,4.8$ & 328.9885 & $-$14.6079 & 189 & 52 & 3 \\
Alessi 3 & 109.2165 & $-$46.3786 & $275.2\,\pm\,2.6$ & 257.6573 & $-$15.1952 & 88 & 47 & 1 \\
Aveni-Hunter 1 & 354.3071 & 48.3612 & $416.4\,\pm\,5.6$ & 110.4164 & $-$12.6989 & 113 & 29 & 2 \\
Blanco 1 & 0.8825 & $-$29.9150 & $234.3\,\pm\,0.8$ & 15.2846 & $-$79.1200 & 519 & 117 & 1 \\
Collinder 135 & 109.3361 & $-$36.7609 & $293.9\,\pm\,1.8$ & 248.7192 & $-$11.0970 & 212 & 66 & 1 \\
Collinder 350 & 267.0184 & 1.4074 & $366.0\,\pm\,3.3$ & 26.8441 & 14.7183 & 155 & 76 & 1 \\
Gulliver 20 & 273.6913 & 11.1062 & $419.8\,\pm\,5.0$ & 38.8340 & 13.1195 & 80 & 28 & 3 \\
IC 2391 & 130.2486 & $-$53.0266 & $150.5\,\pm\,0.5$ & 270.3989 & $-$6.7794 & 198 & 69 & 3 \\
IC 2602 & 160.7801 & $-$64.3706 & $150.4\,\pm\,0.7$ & 289.6021 & $-$4.8724 & 249 & 88 & 3 \\
IC 4665 & 266.3973 & 5.6721 & $340.9\,\pm\,3.1$ & 30.4945 & 17.2198 & 179 & 52 & 1 \\
Mamajek 1 & 130.6553 & $-$78.3476 & $100.9\,\pm\,1.8$ & 291.8782 & $-$21.2700 & 15 & 10 & 1 \\
Mamajek 2 & 264.3584 & $-$7.9914 & $195.1\,\pm\,2.3$ & 17.0597 & 12.4905 & 62 & 20 & 1 \\
Mamajek 3 & 81.8939 & 6.6251 & $98.5\,\pm\,2.3$ & 197.0021 & $-$15.2903 & 24 & 10 & 1 \\
Mamajek 4 & 276.5463 & $-$50.9501 & $441.2\,\pm\,3.1$ & 343.8369 & $-$17.0053 & 90 & 48 & 1 \\
Melotte 111 & 185.8656 & 26.0623 & $85.5\,\pm\,0.6$ & 221.0263 & 83.6569 & 99 & 62 & 1 \\
Melotte 20 & 51.5358 & 48.9839 & $173.3\,\pm\,0.5$ & 147.3072 & $-$6.4265 & 523 & 173 & 5 \\
Melotte 22 & 56.6517 & 24.0765 & $135.0\,\pm\,0.3$ & 166.5277 & $-$23.6113 & 996 & 358 & 1 \\
Melotte 25 & 66.9682 & 16.5421 & $46.5\,\pm\,0.1$ & 179.6565 & $-$21.7361 & 101 & 73 & 2 \\
NGC 1039 & 40.5334 & 42.7536 & $491.6\,\pm\,1.9$ & 143.6702 & $-$15.6167 & 580 & 219 & 2 \\
NGC 1662 & 72.1526 & 10.8962 & $402.6\,\pm\,2.6$ & 187.7547 & $-$21.1051 & 233 & 97 & 1 \\
NGC 1901 & 79.8301 & $-$68.3214 & $415.2\,\pm\,4.1$ & 278.8643 & $-$33.5280 & 74 & 42 & 1 \\
NGC 2516 & 119.5561 & $-$60.7628 & $407.3\,\pm\,0.6$ & 273.8338 & $-$15.8432 & 2211 & 626 & 1 \\
NGC 2632 & 129.9961 & 19.6375 & $183.4\,\pm\,0.4$ & 205.9118 & 32.3824 & 732 & 324 & 1 \\
Platais 3 & 68.8478 & 71.2787 & $178.0\,\pm\,1.9$ & 138.9824 & 15.7877 & 69 & 32 & 1 \\
Platais 4 & 77.1983 & 22.3997 & $313.5\,\pm\,26.9$ & 180.8842 & $-$10.5194 & 74 & 20 & 1 \\
Stephenson 1 & 283.5958 & 36.8612 & $354.6\,\pm\,2.0$ & 66.8801 & 15.3303 & 321 & 63 & 1 \\
Turner 5 & 142.9664 & $-$36.4349 & $416.7\,\pm\,3.9$ & 264.1163 & 10.9797 & 41 & 22 & 1 \\
UBC 1 & 287.9726 & 56.8465 & $323.2\,\pm\,1.8$ & 87.5620 & 19.7780 & 116 & 43 & 4 \\
UBC 12 & 126.0640 & $-$8.5525 & $437.6\,\pm\,9.4$ & 231.7665 & 16.1980 & 24 & 16 & 4 \\
UBC 159 & 339.0372 & 39.6582 & $489.7\,\pm\,2.7$ & 96.4360 & $-$16.1532 & 78 & 23 & 4 \\
UBC 31 & 61.0929 & 32.8044 & $356.2\,\pm\,1.6$ & 163.2998 & $-$14.5508 & 487 & 110 & 3 \\
UBC 7 & 106.6258 & $-$37.6535 & $273.9\,\pm\,1.7$ & 248.6182 & $-$13.4307 & 167 & 34 & 4 \\
UBC 8 & 84.4349 & 57.1315 & $471.9\,\pm\,2.7$ & 154.8780 & 13.3166 & 144 & 60 & 4 \\
UBC 9 & 276.7157 & 26.4682 & $343.7\,\pm\,3.2$ & 54.5648 & 16.7681 & 106 & 32 & 4 \\
UPK 303 & 39.3939 & 40.6252 & $211.0\,\pm\,2.2$ & 143.8065 & $-$17.9051 & 25 & 7 & 4 \\
UPK 305 & 40.3849 & 38.8866 & $400.1\,\pm\,10.2$ & 145.3155 & $-$19.1525 & 39 & 13 & 3 \\
UPK 350 & 101.7852 & 48.6801 & $426.8\,\pm\,4.4$ & 167.3346 & 19.3564 & 60 & 16 & 3 \\
UPK 385 & 75.5438 & 12.9241 & $313.0\,\pm\,5.7$ & 187.9761 & $-$17.2222 & 57 & 11 & 3 \\
UPK 533 & 123.2414 & $-$52.4858 & $336.6\,\pm\,3.0$ & 267.5120 & $-$9.9585 & 62 & 19 & 3 \\
UPK 540 & 114.8198 & $-$58.6372 & $362.2\,\pm\,3.1$ & 270.7512 & $-$17.0598 & 85 & 19 & 3 \\
UPK 606 & 215.7553 & $-$46.4112 & $168.3\,\pm\,2.1$ & 318.9521 & 13.5922 & 36 & 19 & 3 \\
UPK 612 & 254.1704 & $-$68.3064 & $225.1\,\pm\,1.5$ & 322.5872 & $-$15.4121 & 74 & 25 & 3 \\
UPK 624 & 254.3194 & $-$60.5082 & $311.8\,\pm\,4.0$ & 329.0544 & $-$10.8109 & 39 & 14 & 3 \\
UPK 654 & 277.2507 & $-$34.5111 & $496.0\,\pm\,6.3$ & 359.5211 & $-$10.7585 & 52 & 13 & 3 \\
UPK 88 & 309.8308 & 21.1573 & $285.2\,\pm\,4.1$ & 64.6697 & $-$12.2315 & 32 & 8 & 4 \\
 \hline
\end{tabular}
\tablefoot{
N(init) and N(RV) are the number of stars in the initial sample (with radial velocities, respectively).}
\tablebib{(1) \citet{2013A&A...558A..53K}, (2) WEBDA, (3) \citet{2021MNRAS.504..356D}, (4) \citet{2021A&A...647A..19T}, (5) Simbad.}
\label{tab.selected_clusters}
\end{table*}

\subsection{Stellar positions and velocities} \label{sec.positions_velocities}
We downloaded the catalogue with astrometric measurements of the Milky Way stars from the \textit{Gaia}\,DR3 database \citep{2016A&A...595A...1G, 2022arXiv220800211G}.
We limited the selection to stars with parallax uncertainties better than 20\% but applied no constraints on any other parameters except for parallax to encompass a parameter space large enough to include all possible candidate members for all clusters. 
Due to the fact that stars with \texttt{1/parallax} just beyond 500 pc have relatively large parallax uncertainties (typical \texttt{parallax\_error} of 12\% as opposed to 9\% for stars between 400 and 500\,pc and 4\% for stars between 100 and 200\,pc) 
and can thus still be consistent with the distance of a cluster at 500 pc, we set the parallax limit to 1.6\,mas, which corresponds to distances up to 625 pc. Our input catalogue therefore includes right ascension, declination, parallax, proper motions, and associated uncertainties for 44\,million 
stars. 

To avoid the minimal risk of negative parallaxes for our most distant stars, parallax uncertainties needed to be taken into account in the distance determination (e.g. see Figure 3 in \citealp{2018A&A...616A...9L}).
Stellar distances were obtained from \citet{2021AJ....161..147B}, who used parallaxes from \textit{Gaia}\,EDR3\footnote{We note that parallaxes in \textit{Gaia}\,EDR3 and \textit{Gaia}\,DR3 are the same.} \citep{2021A&A...649A...1G} with a prior constructed from a 3D model of the Galaxy. In particular, we worked with their geometric distances \texttt{r\_med\_geo} and their lower and upper limits \texttt{r\_lo\_geo} and \texttt{r\_hi\_geo}, respectively. However, their Galactic prior with a scale length varies across the sky. If this scale distance is far away and the parallax uncertainty is high, this may result in a mode distance considerably away from the true one. For parallaxes with an error of 20\%, this translates to a distance error of $10-20$\% for a cluster at 500 pc, as shown in Fig.\ 1 of \citet{2020A&A...644A...7O}. Despite the exclusion of the clusters in the prior of \citet{2021AJ....161..147B}, the effect on the derived distances is negligible. The difference with the inverted parallax distance is minimal, typically less than 1\,pc for stars closer than 400\,pc and 4\,pc for stars beyond this limit.

The latest \textit{Gaia}\,DR3 catalogue \citep{2022arXiv220605902K} contains radial velocities for 33 million stars. 
The overall radial velocity completeness ratio in our sample is 14\%, and 44\% of the stars with radial velocity have uncertainty better than 10\%. The completeness varies over spectral types, from 92\% for G stars to 2\% for M stars, with typical uncertainties of 1.44\,km\,s$^{-1}$ for G and 4.60\,km\,s$^{-1}$ for M stars, respectively.

While the majority of our clusters are composed of relatively faint stars, the nearest ones contain stars visible to the naked eye. Due to the saturation limit at the bright end, 20\% of stars brighter than magnitude three do not have an entry in \textit{Gaia}\,EDR3 \citep{2021A&A...649A...5F}. Additionally, a fraction of bright sources, for example Alcyone, the brightest star in the Pleiades cluster, lack measurements of parallaxes and proper motions. 
Fortunately, the saturation limit in \textit{Gaia}'s predecessor, the \textit{Hipparcos} space telescope, was much brighter. We therefore took parallaxes, proper motions, and their uncertainties from the new reduction of the \textit{Hipparcos} measurements \citep{2007A&A...474..653V}. We used the cross-match with \textit{Gaia}\,DR3 that is available in the \textit{Gaia} archives and added the measurements to our list in the following way: If a \textit{Gaia} source brighter than $G$\,=\,6 mag lacked parallax, we inserted the \textit{Hipparcos} data for this star. Furthermore, there is a small number of \textit{Hipparcos} stars that do not have a counterpart in \textit{Gaia}. We appended such stars brighter than $H_{p}$\,=\,6 mag to our list and noted the underlying assumption of a perfect cross-match. This might not be true in case of binary stars unresolved in \textit{Hipparcos}.
We checked the offsets of the Hipparcos catalogue with respect to \textit{Gaia} for the entire overlapping sample (100,000 stars) and provided the median values. There is an offset of 0.1087 \,mas in the parallax. This is below the typical parallax uncertainty of 0.3600\,mas in Hipparcos but greater than a typical \textit{Gaia} parallax error of 0.0662\,mas in this magnitude range. The offsets of proper motions are 0.0972 and $-$0.0165\,$\mathrm{mas\,yr^{-1}}$ in right ascension and declination, respectively. These values are not only smaller than Hipparcos uncertainties (0.3400 and 0.2900\,$\mathrm{mas\,yr^{-1}}$, respectively) but also smaller or comparable with the \textit{Gaia} uncertainties (0.0658 and 0.0606\,$\mathrm{mas\,yr^{-1}}$, respectively).

\subsection{Stellar masses} \label{sec.mass}
One of the steps in our membership estimation algorithm described below in Section\,\ref{sec.algorithm} is the determination of the systemic velocity for each cluster (i.e.\ the mean velocity of the centre of mass). 
This means that we needed to estimate individual stellar masses. Approximate mass computation is sufficient for our needs, assuming a spherical symmetry of the clusters and the number of members large enough to provide a robust estimate.
This presumption is supported by \citet{1998A&A...331...81P} and \citet{2018MNRAS.477.3197R}, who concluded that the chosen weighting scheme in mass determination does not have a large influence on the resulting barycentre. 
Similarly, assuming an isotropic distribution of binaries in clusters, their mass determination technique does not have an impact on the barycentre \citep{2019A&A...623A..35L}. Unresolved binary stars in our catalogue thus have masses determined in the same way as single objects. We note that this does, however, skew the total mass of the cluster.

\begin{figure}
\includegraphics[width=\linewidth]{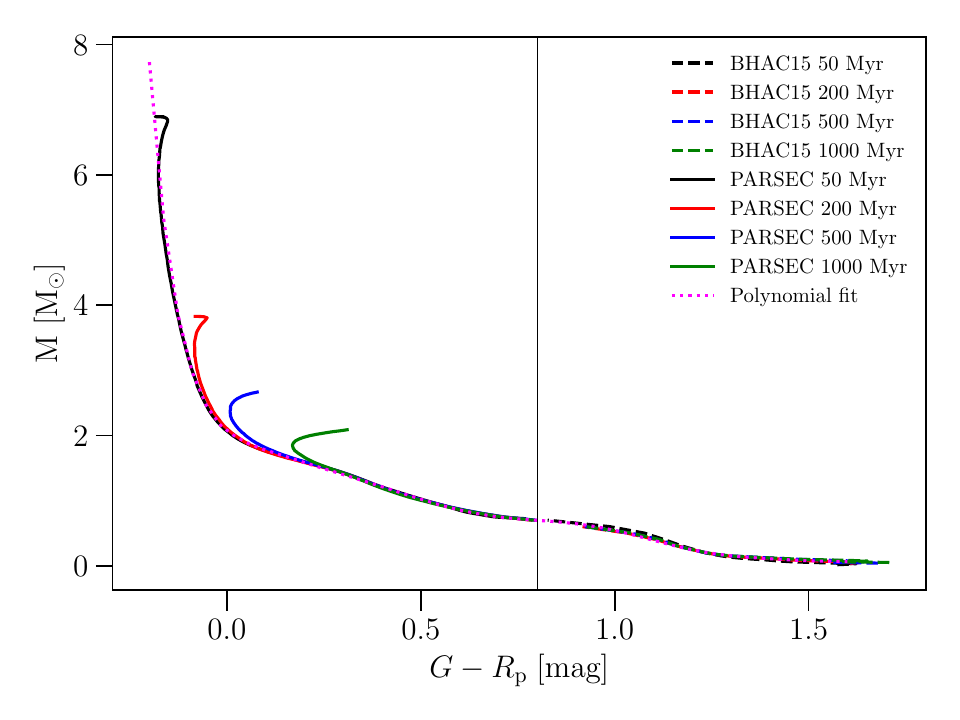} 
\caption{Comparison of isochrones for a range of ages from 50\,Myr to 1\,Gyr for the BHAC15 (dashed lines) and PARSEC models (solid lines). 
The polynomial fit used as a colour-mass relation is shown in magenta. While we only plot the relation for $G-R_\mathrm{P}>-0.2$, the fit extends to bluer colours in order to cover the most massive stars in our input catalogue.
}
\label{fig.isochrones}
\end{figure}

Mass estimation using the luminosity-mass or magnitude-mass relations requires a particular model for each cluster to meet its age and metallicity. To simplify the process, we adopted a colour-mass relation that is valid for a wide range of masses and ages below 1\,Gyr. A comparison of different isochrones with ages between 50\,Myr and 1\,Gyr in Figure\,\ref{fig.isochrones} shows no large variation that would exceed our requirements of rough mass estimates. We thus chose the models at 200\,Myr.

Our mass estimation is based on the colour-mass relation from two models that are joined together at the point of their best overlap at $G-R_\text{P}$\,=\,0.8 mag. 
The vast majority of cluster member candidates have positive $G-R_\text{P}$ colours.
We took the isochrones with solar metallicity from \citet{2015A&A...577A..42B} for $G-R_\text{P}$\,$>$0.8 mag and PARSEC isochrones (the PAdova and TRieste Stellar Evolution Code; \citealp{2012MNRAS.427..127B, 2014MNRAS.444.2525C, 2015MNRAS.452.1068C, 2014MNRAS.445.4287T, 2017ApJ...835...77M, 2019MNRAS.485.5666P, 2020MNRAS.498.3283P})
for stars with for $G-R_\text{P}$\,$<$\,0.8 mag. 
We fit an 11th order polynomial to these two isochrones to provide a relation that can be applied to all stars in our list. To avoid the extrapolation problems outside the models, we set the 
lower mass limit to 0.05\,$M_\odot$. 
For dwarf stars with $G-R_\mathrm{p}<0$, we followed a similar procedure but took the 50\,Myr isochrones.

The masses of \textit{Hipparcos} stars are based in the $B-V$-mass relation from an updated table\footnote{\url{http://www.pas.rochester.edu/~emamajek/EEM_dwarf_UBVIJHK_colors_Teff.txt}, v2021.03.02.} originally published by \citet[hereafter we cite the table as \citetalias{2013ApJS..208....9P}]{2013ApJS..208....9P}. We fit a 21st order polynomial to this relation and estimated the masses of \textit{Hipparcos} stars using their $B-V$ colours. 
Again, we reinforced all masses to be greater than 0.05\,$M_\odot$. 
We set masses for a few individual stars above 5\,$M_\odot$ manually from \citet{2011MNRAS.410..190T} in mass determination with both colours.

Some of our older clusters contain white dwarfs. Typically, about 4\% of objects in the input catalogue are found in the white dwarf region of the colour-magnitude diagrams. We defined all stars fainter than $M_\mathrm{G}=10(G-R_\mathrm{p})+5$ as white dwarfs and assigned them a typical mass of a DA white dwarf (hydrogen lines), 0.59\,M$_\odot$ \citep{2007MNRAS.375.1315K}, which represent 80\% of all white dwarfs. 
This sample partially overlaps with 
a list of white dwarfs identified in \textit{Gaia}\,EDR3 \citep{2021MNRAS.508.3877G}. 

\subsection{Field population} \label{sec.field}
We compared the luminosity functions and mass distributions of the clusters with the field population. 
We thus prepared a sample of field stars in the following way: We selected stars from the \textit{Gaia}\,DR3 catalogue within 25 pc with small parallax errors by requiring \texttt{parallax\,$>$\,40}\,mas and \texttt{parallax\_error\,$<$\,8}\,mas in the \textit{Gaia} query. We excluded stars with either missing or noisy photometry in the $G$ and $R_{p}$ filters. In particular, we required \texttt{phot\_rp\_mean\_flux\_over\_error} and \texttt{phot\_g\_mean\_flux\_over\_error} to be more than 100.
These selection criteria resulted in 4760 stars that we proclaim as the field population. 
We computed the total mass within the given distance to the Sun. This cumulative mass distribution -- the total mass within the given distance to the Sun -- reflects a uniform distribution with no over-densities. We tested this claim by taking the same sample and simulating a uniform distribution of the stars within a sphere with the same volume. In practice, we kept the original locations of the stars but shuffled their masses. The cumulative mass distributions of both samples are practically the same and correspond to a mass density of 0.033\,$\mathrm{M_\odot pc^{-3}}$ (see Figure\,\ref{fig.tidal_radius_pleiades}).  
This sample was assumed to be complete down to L4 objects, while the completeness for the most distant clusters in our list does not exceed mid-M dwarfs. For comparison, we expected it to be complete to M9/L0 in the Hyades (see our reasoning in Section \ref{sec.mass}) and $\sim$M7 in the Pleiades \citep{2019A&A...628A..66L,2019A&A...623A..35L}. This means that membership lists for most of the clusters in this work lack brown dwarfs.
We note that the comparison with the field is not ideal for more distant clusters, since these samples might not come from the same environment and are affected by  completeness and contamination issues.

\subsection{Photometric data} \label{sec.photometry}
We cross-matched our catalogues of cluster members (see Section\,\ref{sec.catalogs_description}) with the photometric surveys described in Table \ref{tab.photometric_surveys} to complement the \textit{Gaia} astrometry and photometry of all the stars.
We used cross-matches provided in the \textit{Gaia} database for 2MASS, AllWISE, and Pan-STARRS.
For SDSS, UKIDSS, and VISTA, we performed a coordinate cross-match in TopCat within the radius of 3\,arcsec.

\begin{table*}
\caption{Photometric surveys to complement \textit{Gaia} data.}
\begin{tabular}{l l l}
\hline
Photometric survey & Photometric bands & Reference \\
\hline
Two Micron All-Sky Survey (2MASS) & $J$, $H$, $K_{s}$ & \citet{cutri03,2006AJ....131.1163S} \\
Wide Field All Sky Explorer (AllWISE) & $W1$, $W2$, $W3$, $W4$ & \citep{wright10} \\
UKIRT Infrared Deep Sky Survey (UKIDSS) & $Y$, $J$, $H$, $K$ & \citep{lawrence07} \\
VISTA Hemisphere Survey (VHS) & $Y$, $J$, $H$, $K_{s}$ & \citep{mcmahon12} \\
Sloan Digital Sky Survey (SDSS) DR13 & $u$, $g$, $r$, $i$, $z$ & \citep{york00,gunn06a,2017ApJS..233...25A} \\
\multirow{2}{20em}{Panoramic Survey Telescope and Rapid Response System (Pan-STARRS) DR1} &&\\
& $u$, $g$, $r$, $i$, $z$ & \citep{2016arXiv161205560C,2020ApJS..251....7F} \\
\hline
\end{tabular}
\label{tab.photometric_surveys}
\end{table*}

\section{Membership determination} \label{sec.memberships}
The membership determination technique used in this work exploits the fact that stars in open clusters display a common location and velocities, as opposed to the uniform distribution of field stars in a Cartesian parameter space.
The method takes advantage of the full 6D space (i.e.\ stellar positions and velocities) and the 5D space, when radial velocities are missing. It is based on the comparison of the transverse and radial velocities of a candidate star with the values expected for a member of a cluster at the location of a star. The expected values were estimated from the cluster's systemic velocity and the position of the candidate star with respect to the cluster. 
This approach was developed by \citet{1998A&A...331...81P} to retrieve the members of the Hyades cluster from the \textit{Hipparcos} data with a later application to Gaia DR1 \citep{2018MNRAS.477.3197R}. 
\citet{2019A&A...623A..35L} and \citet{2019A&A...628A..66L} utilised the same method on the \textit{Gaia}\,DR2 data to improve the memberships of the Hyades, {\ensuremath{\alpha}}\,Persei, the Pleiades, and Praesepe clusters. The specifics of this technique are described in the aforementioned papers. 
In this section, we describe the input data for each cluster and follow it with a summary of our membership selection algorithm, which is an improved version of the one used by \citet{2019A&A...623A..35L}.

The terms we use to describe different samples of stars in the next sections are the following: 
`Preliminary members' are used only in the membership determination method but are not part of the results; see \ref{sec.preliminary_members}. 
`Bona fide members' are the most reliable members of the cluster, as these are stars within 1 tidal radius of the cluster centre (see \ref{sec.clusters_and_their_members}). Finally, `cluster members' include stars up to 3 tidal radii.

\subsection{Input catalogues for each cluster}
The input catalogues for each cluster consist of two parts, the list of preliminary candidate members and the list of objects in the cluster region. The purpose of the catalogue with the preliminary candidate members is the computation of the initial systemic velocity and barycentre of the cluster, which were later refined in the iterative loop described in Section\,\ref{sec.algorithm}. 
As for the list of objects in the cluster region, it encompasses a large volume around the cluster in order to include all potential candidate members. This list is a simple subset of our entire \textit{Gaia} catalogue with 44\,million stars, outlined in \,Sect. \ref{sec.positions_velocities}, and its purpose is to speed up the membership selection algorithm.
We added Cartesian coordinates for stellar velocities and positions as required in some steps of the membership determination algorithm. 
In the following subsections, we describe the preparation process for both datasets and the coordinate transformations.

\subsubsection{Preliminary candidate members} \label{sec.preliminary_members}
The catalogue of preliminary candidate members was prepared manually for each cluster to ensure a coherent catalogue. 
The goal was to find enough bona fide members to estimate the initial systemic velocity and barycentre of a cluster in a robust way. This means that our goal was to have a low contamination rate and that we were not concerned with the completeness level. 

In the first step, we limited the data to the region of the sky centred on a cluster with coordinates reported in the literature. This region typically spans an angle in the sky that corresponds to a diameter of 60 pc at the distance of the cluster centre. Since the typical radius of a bound cluster is less than 10 pc, this diameter should include all bona fide members. 
We note that not all clusters appear as clear over-densities in the sky. Therefore, we identified each of them as an overdensity in the proper motion space with the help of the literature values. We noted an occasionally complex and crowded nature for the proper motion space with asymmetric and sometimes partially overlapping over-densities, due to the rather large selected area in the sky. We refer the reader to Appendix\,\ref{sec.individual_clusters} for a description of clusters with peculiarities.
The cut in the proper motion space is circular with a typical radius of 0.7\,mas\,yr$^{-1}$. We manually adjusted this radius after a visual inspection to cover the obvious overdensity but stayed conservative in order to limit the contamination to the minimum. 

To further help eliminate interlopers, we removed outliers in the distance versus total proper motion space. 
This step revealed a clear, narrow sequence in the colour-magnitude diagram that represents a coeval population of stars.

The selection of initial candidate members was an iterative process because some individual and less populated clusters appeared as somewhat sparse groups of stars in the sky with only a small number of members. We modified the parameters of the cuts as needed to optimise the selection. Finally, we manually removed the outliers in the colour-magnitude diagram, taking out all white dwarfs and (sub)giants as well as any stars with radial velocities significantly different from the average value in the sample (e.g.\ more than 20\,km\,s$^{-1}$). 
This left us with a sample of stars that are very likely members of the cluster. 
The number of initial candidate members and the number of stars with a radial velocity in the initial sample is listed in Table\,\ref{tab.selected_clusters}. The fraction of stars with a radial velocity in the initial sample has increased thanks to \textit{Gaia}\,DR3 and varies from 20--70\%, depending on the spectral type, with a typical value of 30\%. These numbers enable good statistics for determining a robust initial systemic velocity of the cluster.

\subsubsection{Objects in the cluster region} \label{sec.candidate_members}
To ensure the volume of each of the input catalogues was large enough to enclose all potential candidate members, we performed wide and simple cuts to our fully downloaded \textit{Gaia} catalogue (Section\,\ref{sec.positions_velocities}). In particular, we only made cuts in the right ascension and declination in order to limit the catalogue to the cluster centre and its vicinity in the sky. We took the right ascension and declination from the literature and kept all the stars that are located within an angular span and have a radius approximately three times the estimated radius of the cluster. We visually compared the distribution of this input catalogue and the initial list of candidate members in the sky to make sure the volume is large enough. We did not make any additional cuts in the distance in order to mitigate the risk of missing potential members with large distance uncertainties (except for the distance cuts explained in Section\,\ref{sec.positions_velocities}). The typical number of objects in the cluster region varies from a few 10,000 to a few 100,000 stars, depending on the part of the sky and distance from the Galactic plane.

\subsubsection{Coordinate transformations}
We transformed the spatial observables $\alpha$ and $\delta$ as well as the distance $d$ (in parsecs) from the input catalogue of 44\,million stars to the Galactic Cartesian system of physical distances with the centre in the Sun. The position of the $i$th star is described by $\mathbf{b}_i=(b_{xi},\,b_{yi},\,b_{zi})$.
The velocity vector $\mathbf{v}_i=(v_{xi},\,v_{yi},\,v_{zi})$ was computed by the transformation of the observed transverse and radial velocities into the equatorial coordinate frame with
\begin{align}
    \begin{pmatrix}
        v_{xi} \\
        v_{yi} \\
        v_{zi}
    \end{pmatrix}
    =
    \mathbf{R}_i
    \begin{pmatrix}
        &V_{\alpha* i} \\
        &V_{\delta i} \\
        &V_{R i}            
    \end{pmatrix}.
\end{align}
The radial velocity is denoted by $V_{Ri}$. The transverse velocities $V_{\alpha* i}$ and $V_{\delta i}$ are defined as 
\begin{align}
    V_{\alpha*i} &= \mu_{\alpha*i} A_v / \pi_i, \quad \mathrm{and} \label{eq.Valpha} \\
    V_{\delta i} &= \mu_{\delta i} A_v / \pi_i, \label{eq.Vdelta}
\end{align}
where $\mu_{\delta i}$ and $\mu_{\alpha*i}\,=\,\mu_{\alpha i} \cos{\delta_i}$ are proper motions and $\hat{\pi}_i$ is a parallax computed from the distance \texttt{r\_med\_geo}. 
The constant $A_v=4.74047\,\mathrm{km\,yr\,s^{-1}}$ was used to convert the units and express $V_{\alpha*i}$ and $V_{\delta i}$ in kilometres per second.
The rotation matrix $\mathbf{R}_i$ is given by
\begin{align}
    \mathbf{R}_i = 
    \begin{pmatrix}
        -\sin{\alpha_i} & -\sin{\delta_i}\cos{\alpha_i} & \cos{\delta_i}\cos{\alpha_i}\\
        \cos{\alpha_i} & -\sin{\delta_i}\sin{\alpha_i} & \cos{\delta_i}\sin{\alpha_i} \\
        0 & \cos{\delta_i} & \sin{\delta_i}\\
    \end{pmatrix}.
    \label{eq.R}
\end{align}

\subsection{Membership determination algorithm} \label{sec.algorithm}
Stellar clusters appear as over-densities in the spatial and kinematic parameter space.
The process of identifying cluster members thus operates in the full 6D parameter space. The algorithm takes the kinematic properties of the stars in the cluster region and compares them with the systemic velocity of the group. 
However, the members of a cluster must be known in order to determine its systemic properties. 
To break this cyclic dependency, we used a two-step procedure that iteratively improves both the membership list and the systemic cluster properties. In the first iteration, we predefined the systemic cluster properties from the initial list of bona fide members and searched for the candidate members. In the next iteration, the new membership list was used to improve the cluster properties and in turn update the catalogue of candidate members. This iteration was repeated until converged. We proceed next with the details of each iterative step, convergence criterion, and the treatment of stars with missing radial velocities.

\subsubsection{Systemic properties of the cluster}
The systemic properties include the barycentre, systemic velocity, and half-mass radius of the cluster. The computation of the barycentre and velocity was done in the Cartesian coordinate system by
\begin{equation}
    \mathbf{b}_c = \frac{\Sigma m_i \mathbf{b}_i}{\Sigma m_i}, \quad \mathbf{v}_c = \frac{\Sigma m_i \mathbf{v}_i}{\Sigma m_i},
\end{equation}
where $\mathbf{b}_c$ denotes a barycentre and $\mathbf{v}_c$ marks the systemic velocity vector of the cluster. 
The weights $m_i$ are stellar masses (see Section\,\ref{sec.mass}). While the assumption of the entirely single nature of the stars keeps the total cluster mass at its lower limit, it does not affect the barycentre significantly. This is because the contributions of the binary stars cancel out due to their expected isotropic distribution.

We note that radial velocity measurements are not available for all the stars. Therefore, the list of stars used to compute the systemic velocity is a subsample of the catalogue employed in the determination of the barycentre.

The initial cluster systemic velocity and barycentre were determined from the preliminary list of cluster members (Section\,\ref{sec.preliminary_members}). In the later iterative steps, we took candidate members within the 3$\sigma$ confidence interval (see Section\,\ref{sec.step2} for definition) and the half-mass radius of the cluster to refine the systemic cluster parameters.

\subsubsection{Search for cluster members} \label{sec.step2}
Once the systemic cluster properties were determined, we compared the kinematic properties of each object in the input catalogue with those of the cluster. In particular, we computed the expected transverse and radial velocities for a candidate member of a cluster at the location of the candidate star:
\begin{align}
    \begin{pmatrix}
        &V_{\alpha* i}^e \\
        &V_{\delta i}^e \\
        &V_{R i}^e    
    \end{pmatrix}
    =
    \mathbf{R}_i^{-1}
    \begin{pmatrix}
        v_{x} \\
        v_{y} \\
        v_{z}        
    \end{pmatrix}.
\end{align}
The matrix $\mathbf{R}_i$ is defined in Equation\,\ref{eq.R}. The velocity vector in the right-hand side corresponds to the cluster systemic velocity $\mathbf{v}_c=(v_{x},\,v_{y},\,v_{x})$.

To evaluate the membership of a star, we first computed the difference between its observed and expected transverse and radial velocities and denoted it with a vector $\mathbf{z}_i$. To account for the measurement uncertainties and the correlations between the observables, we followed \citet{1998A&A...331...81P} and prepared a covariance matrix for the observed values and another matrix for the expected values of the star. The sum of both matrices $\Sigma$ represents the confidence region and was used to compute the scaled distance between the observed and expected velocity vector by
\begin{equation}
    c = \mathbf{z}^T \Sigma^{-1} \mathbf{z}.
    \label{eq.c}
\end{equation}
Assuming a normal distribution, 
$c^2$ follows an $\chi^2$ distribution for a given number of degrees of freedom. 
For three degrees of freedom (radial velocity and 2D proper motion), a 3$\sigma$ confidence interval corresponds to $c$\,=\,14.16\@. 
For two degrees of freedom, when radial velocity is missing, $c$ equals to 11.83\@.
In other words, stars with a $c$ value equal to or lower than the one corresponding to the selected 3$\sigma$ confidence interval are considered bona fide members of the cluster. 
We note that this part of the algorithm does not take into account over-densities in the positions with respect to the cluster centre.

\subsubsection{Iterations}
The algorithm iterates between the computation of the systemic cluster properties and membership determination until the convergence criterion is achieved. This happens when the absolute difference between the distance of the cluster's barycentre to the Sun in the previous and the new step drops below 0.01 pc, which typically occurred after the second iteration.

Our definition of cluster membership consists of two criteria. The first criterion is that a star must be within a 3$\sigma$ confidence interval in the $c$ value. The second criterion is that only stars that are gravitationally bound to the cluster can be included, that is,\ stars within the tidal radius of the cluster as per definition. 
We describe the determination of the tidal radius in Section\,\ref{sec.tidal_radius}. 

The tidal radius is always larger than the half-mass radius. To make the estimate of the systemic cluster properties more robust, we recomputed the barycentre and cluster velocity with the objects within the tidal radius. 
Finally, we took the new cluster properties and used them to reiterate the tidal radius.

\subsubsection{Tidal radius} \label{sec.tidal_radius}
Cluster members are subject to the gravitational fields of the cluster itself and the Galaxy. In a simplified view, stars with sufficient energy are able to escape the cluster at the point where the gravitational pulls between the cluster and the Galaxy balance out (i.e.\ the first Lagrangian point). We defined the distance of the Lagrangian point from the cluster centre as a tidal radius, assuming that stars are in a circular orbit.
We took Equation 3 from \citet{2011A&A...531A..92R}, which describes the relation between the tidal radius $r$ and the mass of the cluster within the tidal radius $M_\text{cluster}$
\begin{equation}
    M_\text{cluster} = \frac{4 A (A-B)}{G} r^3. \label{eq.tidal_radius}
\end{equation}
Here, $G$ is a gravitational constant, and $A$\,=\,14.5 km\,s$^{-1}$ kpc$^{-1}$ and $B$\,=\,$-$13.0 km\,s$^{-1}$ kpc$^{-1}$ are Oort's constants, from \citet{2006A&A...445..545P}, determined from 581 clusters within 2.5 kpc.

\begin{figure}
\includegraphics[width=\linewidth]{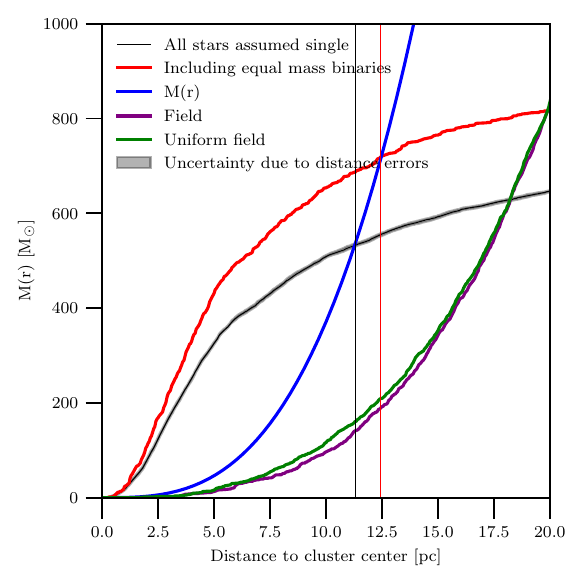} 
\caption{Tidal radius determined by a cross-section between the cumulative radial mass distribution of the cluster and M($r$) from equation\,\ref{eq.tidal_radius} (blue line).
The tidal radius for single stars (black) represents a minimal value, while a synthetic sample where all binaries have an equal mass (red) indicates the upper limit. The difference between these tidal radii is about 10\%. The field distribution was added for comparison (purple and green for the actual and simulated field, respectively; see Section\,\ref{sec.field}).}
\label{fig.tidal_radius_pleiades}
\end{figure}

We solved the equation numerically by comparing the relation from Equation\,\ref{eq.tidal_radius} with the cumulative radial mass distribution of the cluster. 
The intersection of both curves corresponds to the tidal radius. 
The example of the Pleiades is shown in Figure\,\ref{fig.tidal_radius_pleiades}, where we only use stars with a high membership probability (i.e.\ 3$\sigma$ confidence). The figure reveals that the slope of the cumulative radial mass distribution is steeper within the tidal radius, indicating a larger spatial density of the stars. For comparison, we added the distribution of the field stars in the solar neighbourhood (see Section\,\ref{sec.field}) that reflects a uniform distribution of stars with a density of 0.033\,$\mathrm{M_\odot pc^{-3}}$.

\begin{figure*}
\includegraphics[width=\linewidth]{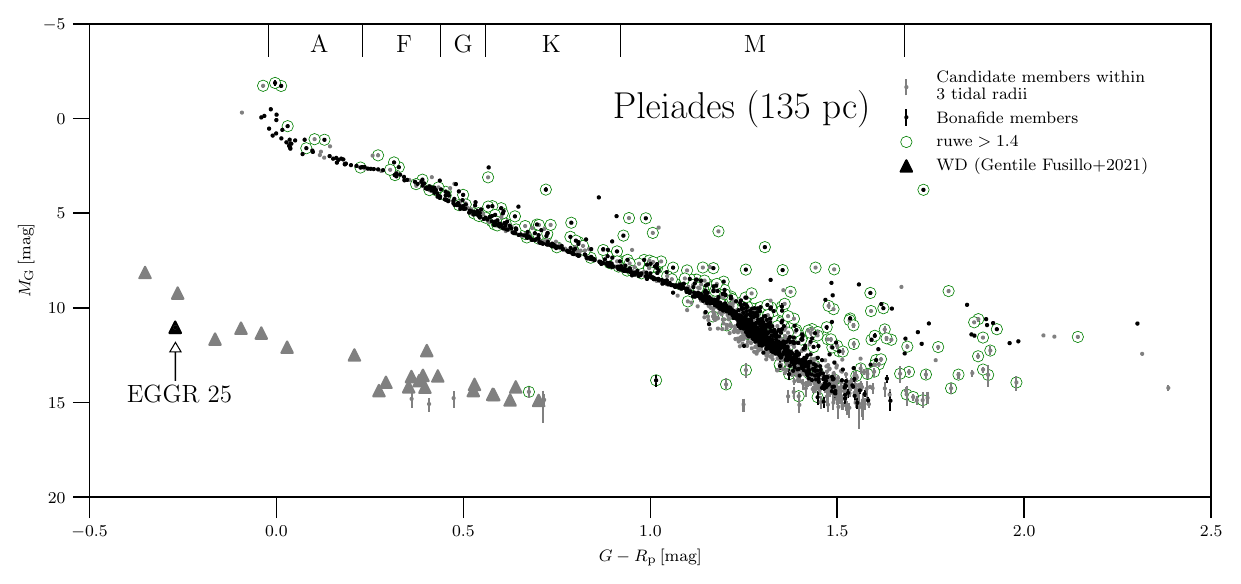} 
\caption{Colour-magnitude diagram for the Pleiades cluster with bona fide and candidate members within 3 tidal radii. The top axis indicates approximate spectral types from the relation in \citetalias{2013ApJS..208....9P}. Stars with high \texttt{ruwe} values (i.e.\ potential multiples) are marked with green circles. Triangles denote white dwarfs from \citet{2021MNRAS.508.3877G}. We found one white dwarf among the bona fide members (EGGR\,25).
}
\label{fig.cmd_pleiades}
\end{figure*}

The cumulative radial mass distribution of the 3$\sigma$ candidate members gives a lower limit of the tidal radius because our mass determination method assumes all stars are single (Section\,\ref{sec.mass}). If we account for binaries, the total mass and thus the tidal radius would be larger. To estimate the upper limit of the tidal radius, we took the members of the Pleiades, manually selected all stars seen by eye above the single cluster sequence (Figure\,\ref{fig.cmd_pleiades}), and proclaimed them as equal mass binaries by doubling their mass. This gave a tidal radius of 12.5 pc that is 11\% larger than the value from the sample with single stars (Figure\,\ref{fig.tidal_radius_pleiades}). However, the true impact of the binaries is likely smaller because not all binaries are equal mass objects. 
Additionally, clusters appear elongated along the line of sight due to the parallax uncertainties, and this affects the determination of their tidal radius. Consequently, not all bound stars are actually included as bona fide members by our selection criteria.
However, we chose to keep this procedure because we only use members within 1 tidal radius to estimate the global cluster properties (e.g. cluster centre in the physical and  velocity space). In our studies of the luminosity functions, we included members up to 3 tidal radii. At such distances to the cluster centre, the stellar density drops, and the distance-to-centre cut does not have such a large dependence on the accuracy of the tidal radius.

\section{Clusters and their members} \label{sec.clusters_and_their_members}
Our membership determination algorithm (Section\,\ref{sec.memberships}) produced catalogues of bona fide members within 1 tidal radius and within a 3$\sigma$ confidence interval in $c$ for all the clusters. We also report candidates within 3 tidal radii for all the clusters in our tables (Section\,\ref{sec.catalogs_description}) in order to be as inclusive as possible and not reject evaporated objects and sources belonging to the corona and tidal tails. 

We divided the list of clusters is into two subsamples (bound and unbound) because tidal radius computation was not possible for the clusters in the unbound sample. Indeed, they appear too sparse and do not reach the minimum total mass that is necessary for a system to be gravitationally bound (and not be disrupted by the Galactic forces).
We manually set their radii to a value that approximately includes most of the obvious members, based on the visual inspection of their overdensity in the proper motion space. We note that their radii are only rough estimates and should be refined for a more detailed analysis.

We further split the bound clusters into three age categories, each of them being represented by one of three benchmark clusters IC\,2602, the Pleiades, and the Hyades.
These three age classes cover different stages in the nuclear and dynamical evolution of the clusters. At the age of $46^{+6}_{-5}\,\mathrm{Myr}$ \citep{2010MNRAS.409.1002D}, IC\,2602  still contains pre-main-sequence stars in the low-mass end. Stars in the Pleiades at the age of 112$\pm$5 Myr 
\citep{2015ApJ...813..108D} have already settled into the main sequence, while the low-mass members of the older Hyades (640$^{+67}_{-49}$ Myr; 
\citealp{2019A&A...623A..35L}) might have already evaporated. The age ranges of the groups are $<$50 Myr for the IC\,2602, 100--200\,Myr for the Pleiades, and $\ge$500\,Myr for the Hyades group. 
The age classification of the clusters into three groups is based on the comparison of their colour-magnitude diagrams with those of the three benchmark clusters. 
We present the results in Table\,\ref{tab.clusters_characteristics}. 
In total, there are seven clusters in the IC\,2602 age group (Alessi\,13, IC\,2391, UBC\,7, Collinder\,135, UBC\,9, and Stephenson\,1), ten clusters similar to the Pleiades ($\alpha$\,Persei, Platais\,3, Blanco\,1, ASCC\,41, ASCC\,101, NGC\,2516, Alessi\,10, Alessi\,24, and NGC\,1039), and six clusters comparable to the Hyades (Coma\,Berenices, NGC\,2632, Alessi\,3, NGC\,1901, and Mamajek\,4).

In Section\,\ref{sec.catalogs_description}, we describe the catalogues and evaluate the membership classification algorithm by comparing our members of the three benchmark clusters with the literature. In Section\,\ref{sec.characteristics}, we outline the systemic cluster properties, provide their coordinates in the position and velocity spaces, and estimate their velocity dispersions. We comment on completeness and contamination rates in Section\,\ref{sec.completeness_contamination}. 

\subsection{Catalogues of cluster members} \label{sec.catalogs_description}
A schema of the membership catalogue is presented in Table\,\ref{tab.catalog_members_format}.  Full membership lists for all clusters are available as supplementary material in the Vizier database\footnote{\url{https://vizier.cds.unistra.fr/}} and on our website, together with the visualisation tools (Section\,\ref{sec.website}). We present all members in one catalogue and note that a small percentage of stars (0.8\%) appear in the catalogue twice. This duplication is due to their membership in close cluster pairs, resulting in their simultaneous assignment to two clusters. (For more details, see Appendix \ref{sec.individual_clusters}.)
We expanded the catalogues with the photometric data from the external databases as outlined in Section\,\ref{sec.photometry}.
Below, we briefly describe the three benchmark clusters, IC\,2602, Pleiades and Hyades, and compare their members from this work with the catalogues from the literature. 
We note that while other works performed more strict data quality cuts --  \citet{2018A&A...618A..93C}, for example, used stars brighter than \textit{Gaia} $G=18$\,mag -- we merely applied filters on the parallax quality and thus reached the detection limit of \textit{Gaia}. Additionally, with the aim of comparing the luminosity functions in mind, we made sure that all our clusters were analysed homogeneously (i.e. all with the same technique). We chose the convergent point method \citep{1998A&A...331...81P} that was designed for astronomy, while most of the literature is based on machine learning techniques that were not originally created for astrophysics.

\subsubsection{Pleiades}
The Pleiades is one of the most prominent and best-studied clusters. Due to its proximity, large number of members, and high proper motion, it stands out from background stars. \citet{2019A&A...628A..66L} provide a comprehensive list of works related to a historical debate about the distance and age of the Pleiades.
The same paper used \textit{Gaia}\,DR2 data to retrieve 1248 sources within the tidal radius of 11.6 pc. The authors estimated a distance of $135.2\pm0.4\,\mathrm{pc}$  
and based its proper motion on 342 stars with radial velocities. They used white dwarfs to estimate the age of the Pleiades to $132^{+26}_{-27} \, \mathrm{Myr}$, in agreement with the lithium depletion boundary age from \citet{1998ApJ...499L.199S, 2004ApJ...614..386B}; and \citet{2015ApJ...813..108D}.

In this work, we found 1355 bona fide members within a tidal radius of $11.28\pm0.03\,\mathrm{pc}$. While our tidal radius is slightly smaller than the value from \citet{2019A&A...628A..66L}, we report about 9\% more members (within the respective tidal radii). This is likely due to more resolved binaries thanks to the longer $Gaia$ baseline.
Our distance of $135.0\pm0.3\,\mathrm{pc}$ 
is in excellent agreement with the aforementioned result.

To make a fair comparison, we took stars within 3\,$\sigma$ in the $c$ value and with a distance to the cluster centre of less than 11.6\,pc in both samples. The overlap between the samples is 1156 stars; 77 stars are found in the \citet{2019A&A...628A..66L} list but are not members in this work. Additionally, this work increases the sample by 223 stars, which is likely due to now-resolved binary stars.

We compared our catalogue with the members from \citet{2020A&A...633A..99C} and found 1045 objects in common. There are 43 stars found only in their catalogue and 334 only in our list. The reason for a larger number of bona fide members in our catalogue is twofold. On one hand, \textit{Gaia}\,DR3 resolved more binaries, and on the other hand, \citet{2020A&A...633A..99C} only considered stars with a five-parameter astrometric solution in the $G$\,=\,5--18 mag range.

The narrow sequence in the colour-magnitude diagram in Figure\,\ref{fig.cmd_pleiades} -- independent of the kinematics used in classification -- reinforces the credibility of the result and is consistent with the coeval nature of the sample typical for clusters. The contamination rate is higher for cooler stars due to their intrinsically low luminosities, binarity, and lack of radial velocity measurements, but the overall contamination rate remains small. Low mass outliers above the main sequence in the $G-R_\mathrm{p}$ colour fall on the main sequence in other colour spaces. 

The binary sequence in the cool part of the colour-magnitude diagram is clearly visible in Figure \ref{fig.cmd_pleiades}. These stars together with other outliers typically have $\mathrm{ruwe}$\,$>$\,1.4, indicating either binarity or problems with the astrometric solution. 

The significant increase in the number of sources with radial velocity measurements in \textit{Gaia}\,DR3 enables a robust statistical study of clusters in the velocity space. 
Except for a few of the brightest stars, practically all stars earlier than $\sim$M4 have a radial velocity measurement. This accounts for about one third of the Pleiades members (480 objects). We inferred a systemic cluster velocity of 32.21$\pm0.27 \,\mathrm{km\,s^{-1}}$ in the Galactic reference frame.

\subsubsection{Hyades}
The nearest star cluster to the Sun is the Hyades, at the distance of 47.03$\pm0.20\,\mathrm{pc}$ \citep{2019A&A...623A..35L}. Its age is estimated to be 640$^{+67}_{-49}\,\mathrm{Myr}$ based on white dwarfs \citep{2019A&A...623A..35L} and 650$\pm70 \mathrm{Myr}$ from its lithium depletion boundary \citep{2018A&A...615L..12L,2018ApJ...856...40M}. 
There are 385 stars within its tidal radius of 9 pc \citep{2019A&A...623A..35L}, but many more are found in its tidal tails. Indeed, \citet{2019A&A...621L...2R} reports 501 members to be within two tidal radii, and 529 stars to be in its asymmetric tidal tails extending up to 170 pc. Similarly, N-body simulation of the Hyades has shown that the total extent of its tails might reach almost 1\,kpc and might thus contain more mass \citep{2021A&A...647A.137J}.

\begin{figure*}
\includegraphics[width=\linewidth]{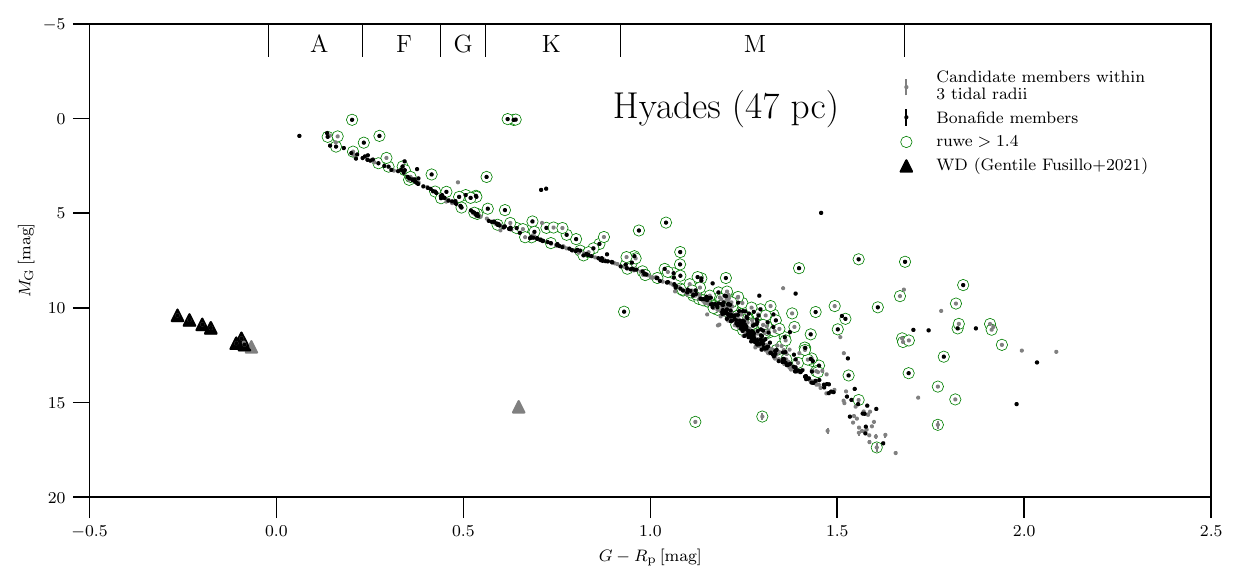} 
\caption{Colour-magnitude diagram for the Hyades cluster with bona fide and candidate members within 3 tidal radii. The top axis indicates approximate spectral types from the relation in \citetalias{2013ApJS..208....9P}. Stars with high \texttt{ruwe} values (i.e.\ potential multiples) are marked with green circles. Triangles denote white dwarfs from \citet{2021MNRAS.508.3877G}. We found seven known white dwarfs among the bona fide members.
}
\label{fig.cmd_hyades}
\end{figure*}

We inferred a distance of 46.47$\pm0.14\,\mathrm{pc}$, which is slightly lower than \citet{2019A&A...623A..35L} but in agreement within 2\,$\sigma$. Our tidal radius (8.4$\pm0.02\,\mathrm{pc}$) is also slightly lower than previous estimates (panel a in Figure\,\ref{fig.tidal_radii}), but the total number of members (427) is slightly larger. 
We note that our input candidate list is centred on $\alpha=67.2149\,\mathrm{deg}$ and $\delta=16.5569\,\mathrm{deg}$ and spans 70$^{\circ}$ in diameter (three times its tidal diameter), while the outskirts of the Hyades extend beyond this limit \citep{2021A&A...647A.137J}. We therefore focussed only on the cluster core and its immediate vicinity. 
Figure\,\ref{fig.cmd_hyades} shows the colour-magnitude diagram of the Hyades and reveals seven white dwarfs that are known members of the cluster.

To compare our sample with the literature, we prepared a catalogue from \citet{2019A&A...623A..35L} by limiting the data to 3\,$\sigma$ in $c$. We took stars with a distance to the cluster centre within 9\,pc in both cases.
There are 346 stars in common between the samples. Additionally, 14 stars were found only in the \citet{2019A&A...623A..35L} catalogue and 108 stars only in this work. We attribute the increase of the sample in this work to a larger fraction of resolved binaries.
We found a similar but slightly smaller overlap in other works. For example, membership lists from \citet{2019A&A...621L...2R} and \citep{2021A&A...647A.137J} respectively have 333 and 318 stars in common with our work. On the other hand, we found 121 and 136 stars from this work that are not in these two catalogues, respectively.

\subsubsection{IC\,2602}
The youngest benchmark cluster in our list is IC\,2602, also known as the Southern Pleiades. Its lithium depletion age is estimated to be 46$^{+6}_{-5} \mathrm{Myr}$ \citep{2010MNRAS.409.1002D} with a distance to 
151.6$^{+1.9}_{-1.8}\mathrm{pc}$ \citep{2022AJ....163..278N}.
We estimated its tidal radius to 7.2 pc in Figure\,\ref{fig.tidal_radii} 
and its distance to 150.39$\pm0.72 \mathrm{pc}$, which are in agreement with the literature.
Its location in the Galactic plane ($b$\,=\,$-5\mathrm{deg}$) makes membership selection difficult. Thus, our membership algorithm results in many interlopers with large photometric uncertainties ($\sigma_G\ge0.15$) at the faint end, which we removed. Since these interlopers have larger relative parallax uncertainties than the more reliable members, we excluded all stars fainter than \textit{Gaia}\, $G=8$\,mag and with \texttt{parallax\_error/parallax}$>$0.075.

Our catalogue contains 308 members within 1 tidal radius and 619 stars within 3 tidal radii, while \citet{2020A&A...633A..99C} and \citet{2022AJ....163..278N}  report 311 and 451 members, respectively; both are based on \textit{Gaia}\,DR2. The overlap between \citet{2022AJ....163..278N} and our sample is 416 stars, while their catalogue contains 44 additional stars that are not in our list.
In our comparison with \citet{2020A&A...633A..99C}, we found 295 objects in common, 324 stars in our list that are not in theirs, and 20 stars in their list that are not in ours. 

\begin{figure*}
\includegraphics[width=\linewidth]{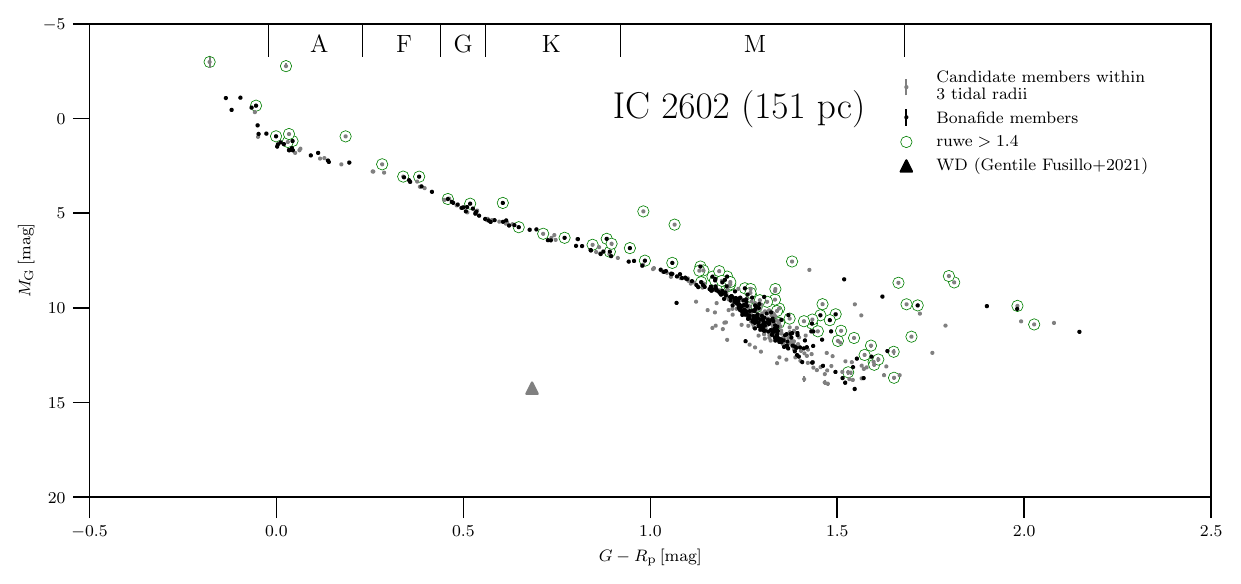} 
\caption{Colour-magnitude diagram for the IC\,2602 cluster with bona fide and candidate members within 3 tidal radii. The top axis indicates approximate spectral types from the relation in \citetalias{2013ApJS..208....9P}. Stars with high \texttt{ruwe} values (i.e.\ potential multiples) are marked with green circles.}
\label{fig.cmd_ic2602}
\end{figure*}

\subsection{Catalogue of cluster characteristics} \label{sec.characteristics}
We present the summary of the cluster properties in Table\,\ref{tab.clusters_characteristics} for bound clusters and Table\,\ref{tab.clusters_characteristics_unbound} for unbound clusters. The tables include the cluster position in the physical and velocity field, expressed in both observable and Cartesian space. We added tidal radius, the number of bona fide cluster members $N$, tidal mass, and velocity dispersion in all three directions.
We computed the velocity dispersions for each cluster. Table\,\ref{tab.clusters_characteristics} lists them separately for transversal and radial velocities due to large uncertainties of the latter. We determined a 2D dispersion $\sigma_{2\mathrm{D}}$ from transversal velocities. 

To check if cluster properties evolve with time, we computed the median values for clusters in each age bin, but we observed no clear trends in the median tidal radius and their dispersion, 
the minimal and maximal tidal radius, nor in 2D dispersion in velocity and the number of bona fide members. This might be due to our age bins being too wide and the fact that while all clusters lose mass with time, the surviving clusters are those with an initially larger mass.


\begin{figure*}
\includegraphics[width=0.9\linewidth]{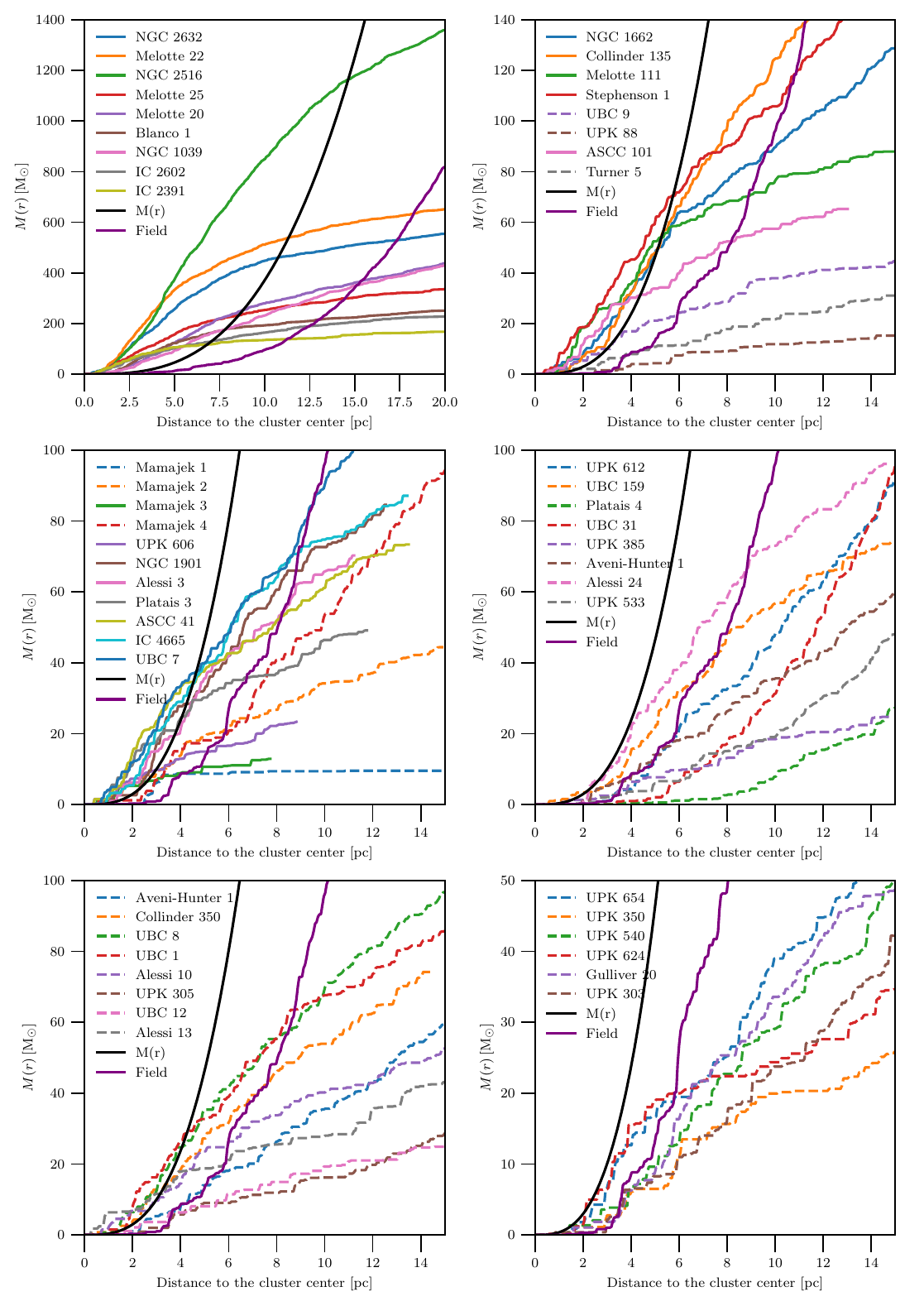} 
\caption{Cumulative radial mass distributions for all clusters. These plots were used to determine tidal radii. The term M($r$) is a theoretical curve that describes the location of the first Lagrange point. The dashed lines represent clusters whose tidal radii we could not determine with our technique, so we set their characteristic radii manually.}
\label{fig.tidal_radii}
\end{figure*}

\subsection{Completeness and contamination rate} \label{sec.completeness_contamination}
In this section, we address the completeness and contamination rate in our catalogues of cluster members. The completeness of the \textit{Gaia}\,DR3 catalogue is affected by the saturation limitations in the bright end and the intrinsic faintness of the low-mass stars. Bright stars missing in \textit{Gaia} are assumed to be covered in our list by the insertion of the \textit{Hipparcos} catalogue, as described in Section\,\ref{sec.data}. 
The completeness ratio at \textit{Gaia} G magnitude 20 is 92.2\% for a five- and six-parameter solution \citep{2021A&A...649A...2L}. Additionally, our input catalogues are affected by the parallax quality cut that correlates with the magnitude and the fact that radial velocities are only available to $G=16\,\mathrm{mag}$.

In \textit{Gaia}, `relatively few sources are found with separations less than about 0.6\,arcsec' \citep{2021A&A...649A...2L}. We checked the crowded regions in the core of the most populated cluster in this work, NGC\,2516, for a possible source of incompleteness. Stellar pairs with a separation of less than 5 arcsec (less than 1\% of the bona fide members) show a rather uniform distribution in the sky. We concluded that stellar density does not affect the incompleteness.

The adopted membership determination algorithm in this work assumes that members have the same motion as the cluster. While the measurement uncertainties in proper motions are small, relatively large errors in parallaxes and radial velocity smear out the internal kinematics, for example, the rotation of the cluster. 
Our estimate is that this algorithm works well for stars within the tidal radius and those that are still found in the vicinity of the cluster and evaporated in the direction of the cluster motion.
However, the technique might not be adequate to identify tidal tails in all directions and large distances from the cluster centre.
Stars that escaped in a direction perpendicular to the cluster motion will not be detected.
Additionally, the results become less trivial to interpret for clusters that have neighbours in both velocity and physical space, for example, Platais\,4 (see Appendix \ref{sec.platais4}). In this case, the algorithm cannot distinguish between the two clusters and detects stars from both clusters as if there was only one large cluster. 

Binary stars orbit their common centre of mass and cause periodic oscillations in their proper motions and radial velocities. Ideally, the velocity of the centre of their mass should be compared with the cluster in our algorithms, but we avoided this step due to the large number of stars in our sample. Instead, we tested the membership detection of binaries in the Pleiades by analysing the distribution of their c values. While \textit{Gaia} resolves wide binary stars with a separation $>0.4\,\mathrm{arcsec}$, \texttt{ruwe}$>$1.4 is a good proxy for close binary objects. Stars with \texttt{ruwe}$>$1.4 have a slightly broader distribution of the c values, with the main peak at c\,=\,1 and a smaller peak at around c\,=\,8. 
This is still well within our 3$\sigma$ confidence interval (c$\approx$14 for three degrees of freedom). Spectroscopic binaries should be detected by \textit{Gaia,} but if the common centre of mass is significantly displaced (e.g. in non-equal mass objects), we may miss them. We concluded that the 3$\sigma$ criterion is good enough to incorporate the majority of binary stars, with a possible exception to the non-equal mass binaries. 

The contamination rate remains small for bound clusters. 
We provide a quantitative estimate from the number of outliers in the colour-magnitude diagram. Contamination in our most distant cluster (NGC 1039 at $b=-15.6\deg$) is up to 2\% within 1 tidal radius and $9-15$\% within 3 tidal radii, depending on how conservative we are in estimating the intrinsic spread in the lower main sequence. 
We note that this is only an approximate estimate, as it is often non-trivial to distinguish between, for example, subdwarfs or overluminous low-mass members and true outliers that might otherwise be reliable kinematic members with problematic photometry.
We automated the contamination estimation procedure and applied it to all clusters in the following way: We fit an 11th order polynomial to the cluster sequence within 1 tidal radius and between $0<G-R_\mathrm{P}<1.5$ and computed a root-mean-square of the sequence. We counted stars that are more than 3\,RMS away from the sequence as contaminants. We note that this is only an approximate estimate, as the polynomial fit is not an ideal solution. We then computed the contamination rate as the number of outliers over the total number of stars in this colour range. 
The contamination rates are below 5\% for most of the clusters.
We found no obvious correlation with the Galactic latitude. Most of our clusters are within 20 degrees from the Galactic plane. Those further away have, on average, contamination rate less than 2\%.
Further, we checked crowdedness in the cluster cores as a possible source of incompleteness and concluded that it is negligible since the pairs with a separation of less than 5 arcsec show a uniform distribution in the sky and are not concentrated in the cluster cores.

\section{The website} \label{sec.website}
Visualisation of the multidimensional parameter space is challenging, especially if the catalogues are large and the phenomena of interest are observed on various scales. To present our results in an adequate way for all the clusters, we designed a website.\footnote{ \url{http://research.iac.es/proyecto/gaiaclusters}} It includes interactive and 3D visualisation tools and allows the user to select the quantities on the axes of the plot for the cluster of interest. The list of available columns includes \textit{Gaia} data and the external photometric measurements described in Section\,\ref{sec.photometry}.
We added a database of PARSEC isochrones to overplot over the colour-magnitude diagrams. 

One of the highlights is the interactive link between our plots and the Aladin sky atlas \citep{2000A&AS..143...33B,2014ASPC..485..277B}\footnote{\url{https://aladin.u-strasbg.fr/}}. In particular, a click on a given star in our plot centres the Aladin plugin on the chosen object. This feature can be useful in many cases, for example, to inspect the astrometric aspect of close binaries and of faint or bright objects{}.
We linked every cluster page with the catalogue of its members to the Vizier database and enabled a direct download of the data for further analysis.

The website was built to enable the professional community to have a quick look at well-known clusters. At the same time, we hope to bring the latest scientific results to the public and inspire new citizen science and exciting student projects.

\section{Luminosity functions} 
\label{sec.luminosity_functions}
The luminosity function represents the number of stars per magnitude bins within a given cluster.
It is subject to a temporal variation due to nuclear and dynamical processes of the stellar population. For example, the low-mass stars in the young cluster IC\,2602 are still found in the pre-main-sequence stage and thus appear more luminous than their older counterparts. In the later phases of a cluster's life, its orbital motion in the Galaxy creates interactions and results in the gradual evaporation of weakly bound and low-mass members. For example, the Hyades lost a significant fraction of M dwarfs from its core. In fact, a significant number of its members very likely drifted away to its tidal tails \citep{2019A&A...621L...2R,2021A&A...647A.137J}. In this work, we also studied the timescale of dynamical evolution and estimated the percentage of M dwarfs lost from clusters at different stages of their life.


To address this question of mass loss, we first re-evaluated the ages of the clusters in the bound subsample from Table\,\ref{tab.clusters_characteristics} whose ages span $\sim$50 to $\sim$650\,Myr in order to cover the most significant period in the evolution of the clusters (Section\,\ref{sec.clusters_and_their_members}). We discuss their constructed spectral type distributions in Section\,\ref{sec.spectral_type_distribution} and their luminosity functions in Section\,\ref{sec.lf}. 
For comparison, we added the field population from the immediate Solar neighbourhood described in Section\,\ref{sec.field} but manually removed white dwarfs because they correspond to objects with larger initial masses.

\subsection{Spectral type distributions} 
\label{sec.spectral_type_distribution}
The low-resolution version of the luminosity function is a spectral type distribution.
We divided stars into spectral types based on their $G-R_{P}$ colour from the \textit{Gaia}\,DR3 catalogue. The spectral type classification is based on \citetalias{2013ApJS..208....9P}. 
To test the reliability of this conversion, we took the sample of field stars within 8 pc from \citet{2012ApJ...753..156K} and compared their spectral types with the classification from the colour conversion. Out of 141 dwarf stars that were cross-matched with the \textit{Gaia}\,DR3 catalogue, 
we found eight cases with discrepant spectral types. 
This is a relatively small error (6\%); 
thus, we concluded that the classification using the $G-R_\mathrm{P}$ colour relation is sufficient for our purpose.

\begin{figure*}
\includegraphics[width=\linewidth]{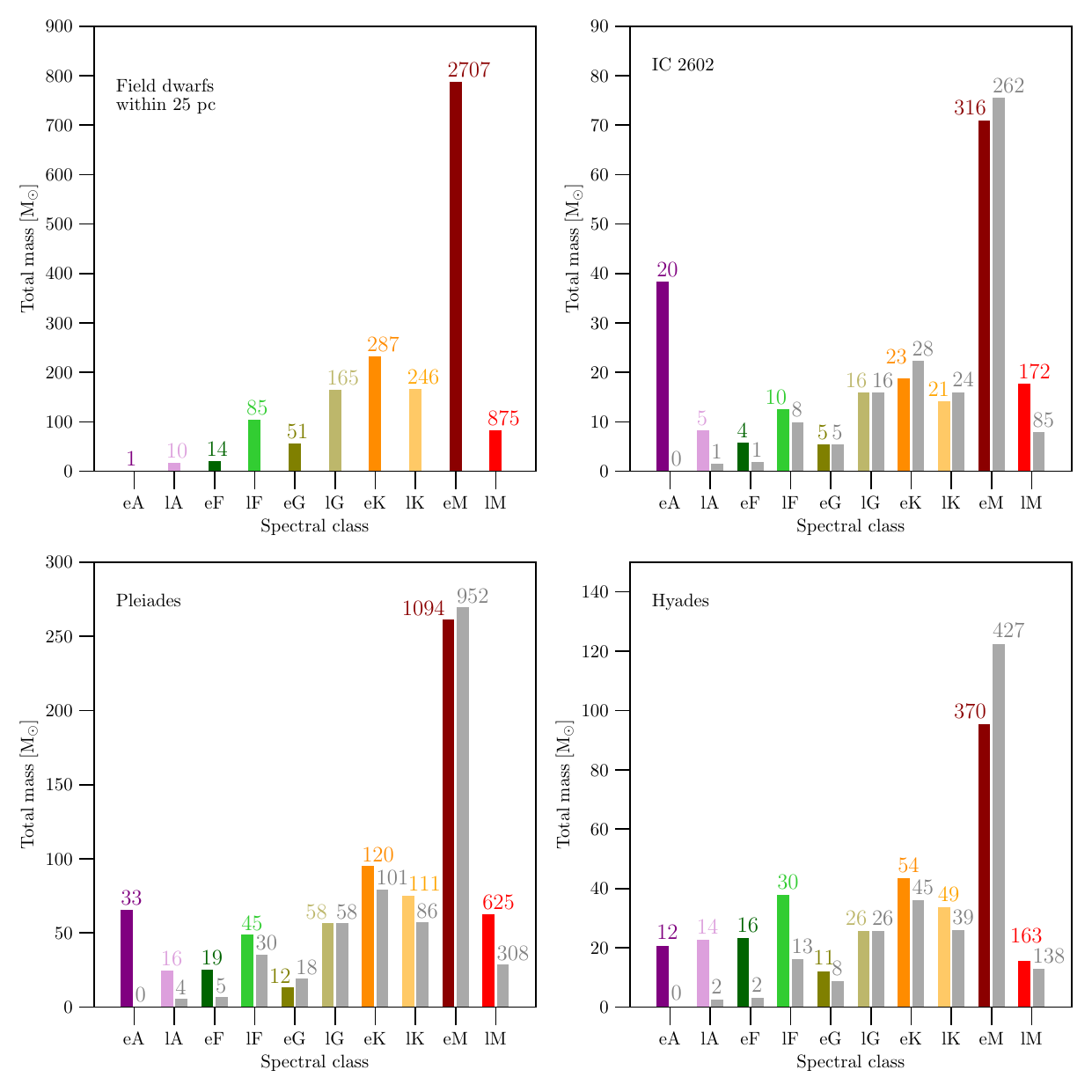} 
\caption{Spectral type and total mass distribution of the field dwarfs within 25 pc and three benchmark clusters, IC\,2602, the Pleiades, and the Hyades. Spectral types G, K, and M are split into early and late subtypes. The field population was used to compute the expected number of stars for each cluster with respect to the number of late-G dwarfs (grey bars).
} 
\label{fig.spectral_type_distribution}
\end{figure*}

Figure\,11 from \citet{2012ApJ...753..156K} presents a sample of all objects within 8 pc in three different ways: the total mass per spectral type, the number of stars per spectral type, and the spatial mass density per spectral type. To reproduce and compare our results with their figure, we took our mass estimates from Section\,\ref{sec.mass} and counted the number of stars per spectral type. We note that binary stars in our work are treated as single objects in the mass determination and thus represent a lower mass limit.
We show the results for our field sample within 25 pc (described in Section\,\ref{sec.field}) and the benchmark clusters in Figure\,\ref{fig.spectral_type_distribution}. In order to compare the clusters with the field population, we overplot the field sample scaled to the total mass and number of late-G stars. To trace the change in the distribution in greater detail, we split the spectral types into early- and late-type classes (i.e.\ from 0--4 and 5--9 subtype bins).

As expected, the number of objects and the total mass grow towards the late spectral types, with a peak around M3--M4 dwarfs. This corresponds to the peak of the present day mass function in the solar neighbourhood and open clusters \citep{kroupa98,2001MNRAS.322..231K,chabrier01,bastian10}. We quantified this observation by computing the fraction of M dwarfs in the entire sample. We did this for masses and the number of stars and present the results in Table\,\ref{tab.ratios}.

The mass and number ratio of M dwarfs in IC\,2602 and the Pleiades match the field population ($\sim$40\% of the total cluster mass and 80\% in terms of a star count). 
In contrast, only 65\% of stars in the Hyades are M dwarfs, representing 30\% of the total cluster mass. Assuming the same initial mass function as for IC\,2602 and the Pleiades, it is clear that the Hyades lost its M dwarfs because of evaporation during its multiple rotations around the Galaxy and the dynamical processes that lead to the formation of the tidal tails that stretch up to nearly 1\,kpc \citep{2021A&A...647A.137J}. The number ratio between the early- and late-M dwarfs in the Hyades is twice as large as in IC\,2602 and the Pleiades. This means that the late-type M dwarfs are missing in the Hyades. Their absence cannot be attributed to \textit{Gaia}'s incompleteness, since the Hyades is the closest of all clusters \citep{lodieu19a}.

\begin{table}
\caption{Number ratios of stars in the field and benchmark clusters.}
\begin{tabular}{l r r r r}
\hline
Cluster & N(M) & N(eM)/N(lM) & N(G) & N(G)/N(M) \\
\hline
Field      & 0.81 & 3.09 & 0.05 & 0.06 \\
IC 2602    & 0.82 & 1.84 & 0.04 & 0.04 \\
Pleiades   & 0.81 & 1.75 & 0.03 & 0.04 \\
Hyades     & 0.72 & 2.27 & 0.05 & 0.07 \\
\hline
\end{tabular}
\tablefoot{N(M) and N(G) are ratios of M and G dwarfs in the entire cluster, while `e' and `l' are early- and late-type M dwarfs (M0--M4, M5--M9). 
}
\label{tab.ratios}
\end{table}

The low-mass stars in the youngest clusters are in their pre-main-sequence phase. Clusters with the Pleiades age can already contain white dwarfs, while clusters in the oldest group with the Hyades age stars can produce (sub)giants. The number of A-type stars in a cluster decreases with age, as expected from stellar evolution models. On the low-mass end, we observed that the numbers of M dwarf members is similar between the field, the Pleiades, and IC\,2602 within the Poisson error bars, while the Hyades shows a clear deficit of low-mass and very low mass stars. We quantified this fact by calculating the ratio of early-M dwarfs to G-type stars in the three benchmark clusters (Table \ref{tab.ratios}). This substantial loss of M dwarfs is the result of interaction during the motion of the cluster in our Galaxy, which yielded the dynamical evaporation of the lower mass members. This appears to be natural in the evolution of a young cluster into an older one, as simulated by \citet{2001MNRAS.321..699K}.

\subsection{Luminosity functions} \label{sec.lf}
A luminosity function is defined as the distribution of the luminosity in a sample (i.e.\ a distribution of stellar magnitudes). Its profile is subject to the intrinsic properties of a population, such as its initial mass function, age, and star formation history, as well as the dynamical processes that cause gradual evaporation of the low-mass stars in the population. 

We determined absolute \textit{Gaia} $G$ magnitudes using the \citet{2021AJ....161..147B} distances 
and assumed a negligible extinction since the clusters are located close to the Sun. 
To check our assumption, we compiled $E(B-V)$ values from the literature for the bound clusters in Appendix \ref{sec.extinction}. Clusters beyond 400\,pc (e.g. Alessi\,10, NGC\,2516, and NGC\,1039) are the most reddened clusters in our sample, with the $E(B-V)$ reaching 0.2\,mag, and we found no correlation with their age. 
Three clusters are missing the literature value, and two of them (UBC\,7: $40-50$\,Myr \citealp{2020A&A...642L...4K}, and UBC\,9, classified as young in this work) are estimated to be very young. We thus provide our own estimate of the upper limit of reddening and extinction based on \texttt{dustmaps} \citep{2018JOSS....3..695M}. The large model reddening values exceed those measured in other works. The upper limit for $E(B-V)$ is 0.21 for UBC\,7, but it is based on other clusters. We deduced that the value is likely much smaller.
Assuming $R_\mathrm{V}=3.1$, the typical extinction $A_\mathrm{V}$ is much less than one magnitude, and we thus concluded that it does not significantly affect our luminosity functions. A possible exception is Alessi\,10, which has an extinction of $A_\mathrm{V}=0.6$ that is approximately half of the magnitude bin in our luminosity function, as described below. 
We computed the luminosity functions in the range from 0 to 15 magnitudes with a fixed step of one magnitude and normalised them by the area under the curve after counting the number of objects per magnitude bin.  

\begin{figure*}
\includegraphics[width=0.33\linewidth]{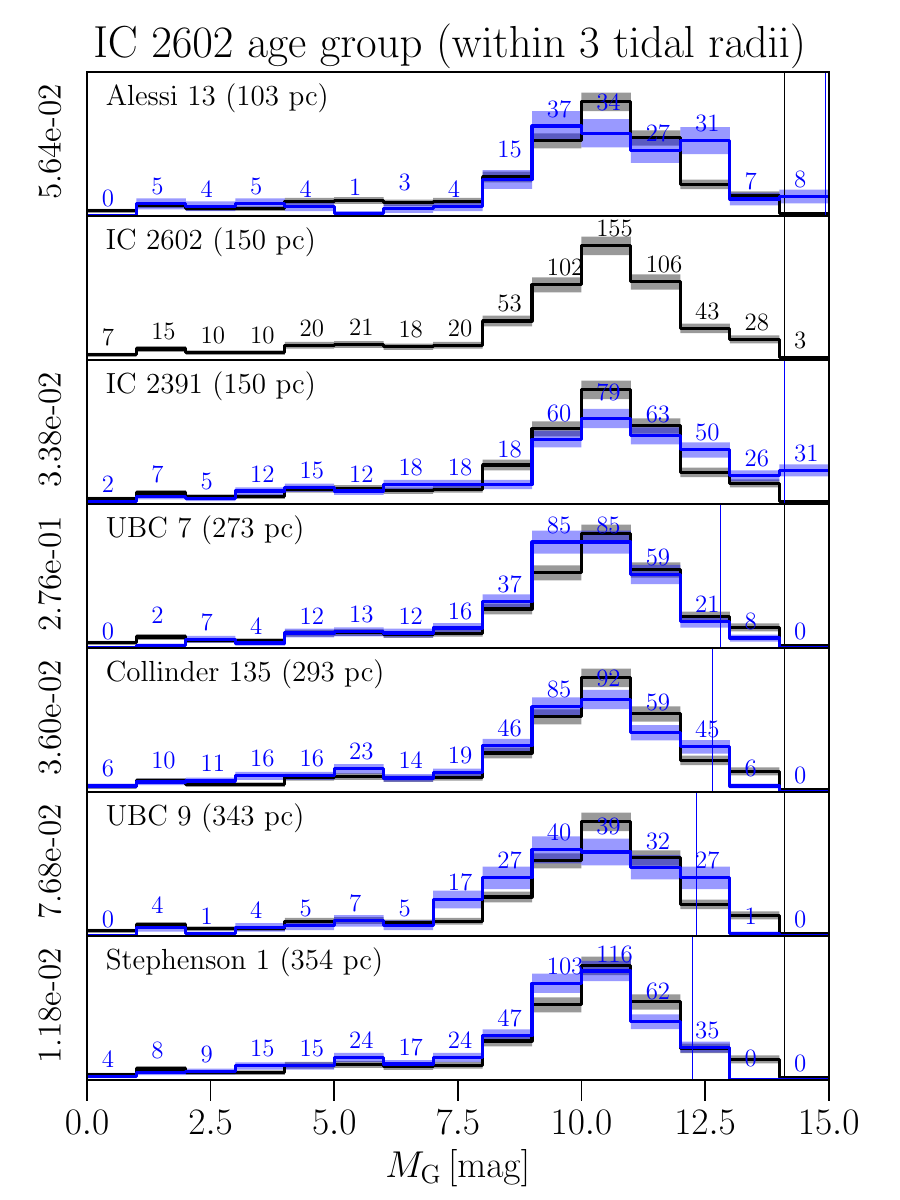}
\includegraphics[width=0.33\linewidth]{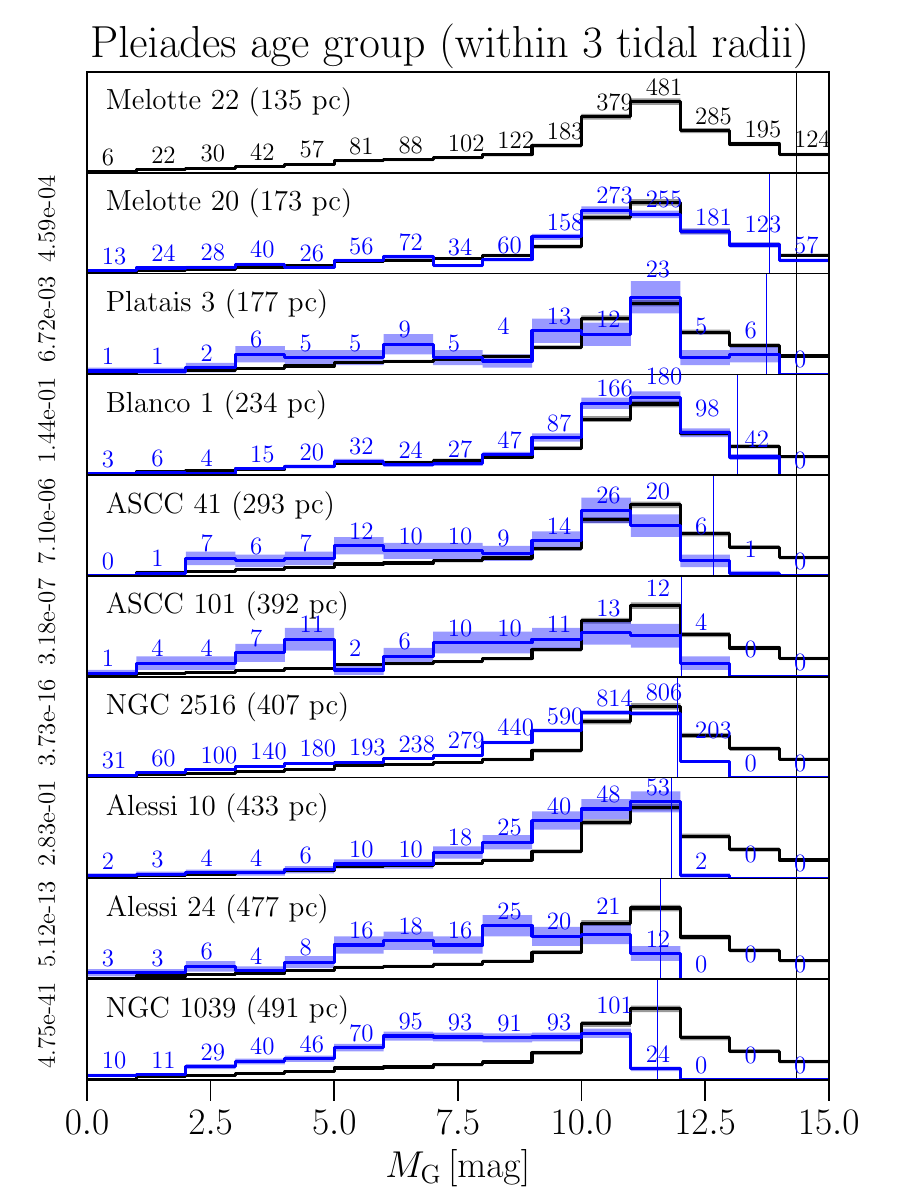} 
\includegraphics[width=0.33\linewidth]{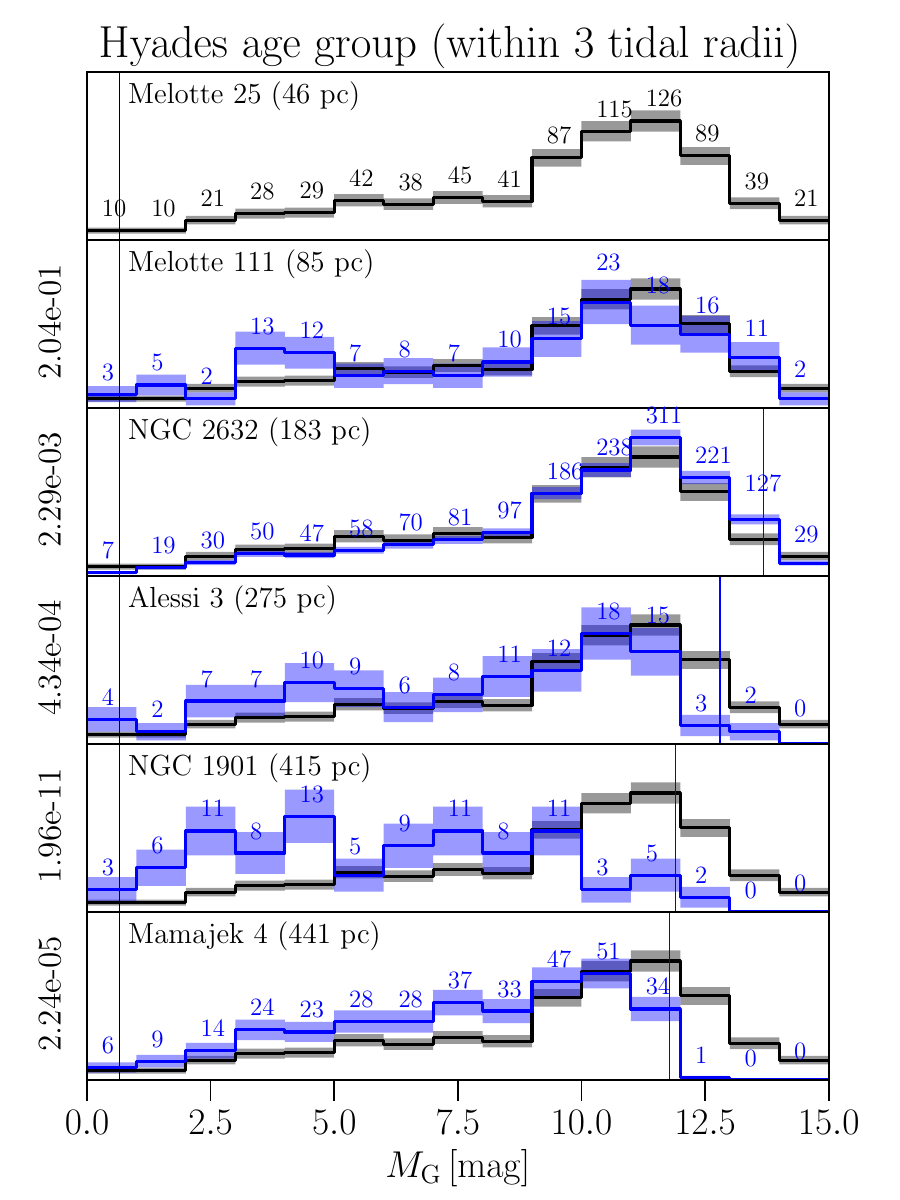} 
\caption{Luminosity functions for clusters in each age group. These functions have been normalised such that the area under the curve equals one. We include all members within 3 tidal radii.
The solid black lines show the benchmark cluster, and shaded areas represent the Poissonian uncertainty. We sort the clusters by their distance to more easily trace the completeness limit (denoted by vertical lines). The y-axes show the p-value from the Kolmogorov-Smirnov comparison, with the benchmark cluster within the common completeness limits.
}
\label{fig.LF_ic2602}
\end{figure*}

\begin{figure*}
\includegraphics[width=\linewidth]{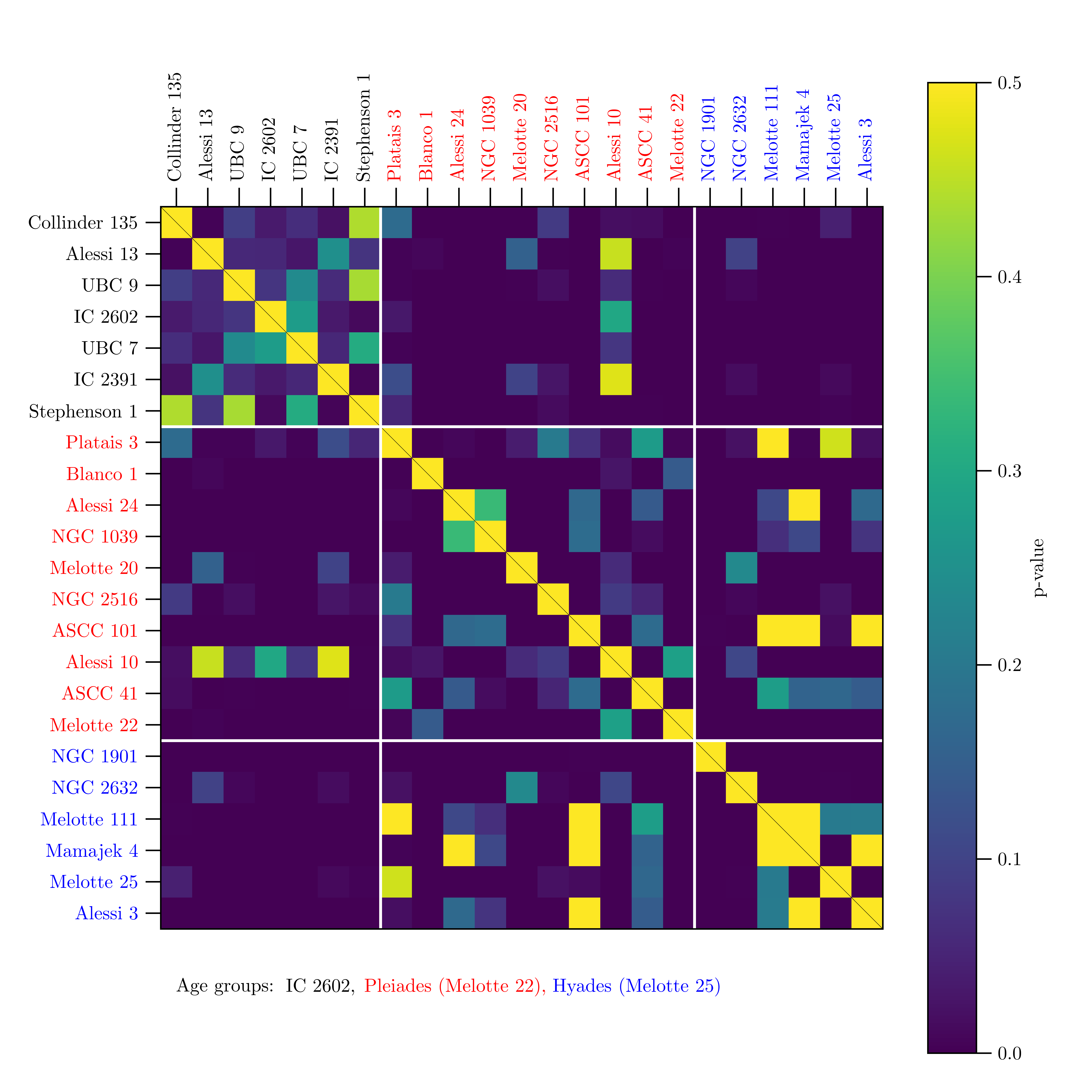} 
\caption{ Pair-wise Kolmogorov-Smirnov test p-values for the luminosity functions of the bound clusters. We sort the clusters by their age group and present them with the following colour code: black for the IC\,2602, red for the Pleiades, and blue for the Hyades age group.
}
\label{fig.tetris}
\end{figure*}

Figure\,\ref{fig.LF_ic2602} shows the luminosity functions of the clusters in each age group together with the benchmark clusters for comparison. In the figure, we plot all members within 3 tidal radii. Within each age group, the luminosity functions seem to be similar to each other, as they practically overlap with the one of the benchmark cluster.
To quantify this comparison, we used a Kolmogorov-Smirnov test for two samples from \texttt{SciPy}. The test was performed pair-wise within the common magnitude completeness limits of each cluster pair. We present the p-values for all pairs in Figure \ref{fig.tetris}, where we sort clusters according to their age group. 
The p-values indicate the probability that the two samples, in our case the luminosity functions of the two clusters in question, come from the same distribution. We note that off the diagonal (the diagonal compares clusters to themselves), the values are generally lower than 0.5, with a few exceptions. 
Because it is not trivial to interpret the absolute values and the test might sometimes be affected by the low-number statistics in the bright end of the luminosity functions, we focussed on the groups as a whole. 
We most often found pairs with a relatively high p-value in the youngest age group. This means that clusters in this age group are more similar to each other than in the intermediate and old age groups. At the same time, it is clear that there is no significant similarity between the young and older clusters from the Pleiades and Hyades age groups, with an exception being cluster Alessi\,10.
On the other hand, there is no significant difference between the Pleiades and Hyades age groups. Clusters NGC\,1901, NGC\,1039, and Platais\,3 are affected by incompleteness and low-number statistics. The average p-values for both age groups, the Pleiades and the Hyades, appear to be lower, and one is less likely to find a pair from the same age group with a high degree of similarity.

Moreover, there are seven clusters that bear similarities to clusters from another age group. These clusters are Alessi\,24, ASCC\,101, and Platais\,3 from the intermediate age group and Melotte\,111, Mamajek\,4, Melotte\,25, and Alessi\,3 from the old age group.

To summarise, clusters in the youngest age group are the most similar to each other, and the degree of similarity later decreases. These observations led us to draw a parallel with the mass functions in star-forming regions, open clusters, and globular clusters, as they are universal within current observational uncertainties (see review by \citealp{2010ARA&A..48..339B}). Based on our data, the luminosity function appears to be universal at least within 500\,pc from the Sun. We reiterate that our cluster selection function only depends on the distance (500\,pc), upper age limit of 1\,Gyr, and the distance from the Galactic plane ($|b|>10$\,deg), and the membership selection bias was minimised by the fact that we used the same method on all of the clusters. Our results are consistent with the earlier study of \citet{2008A&A...487..557P}, who observed a strong similarity between the luminosity functions in nearby clusters and other galaxies, and agree with the findings in young regions, as \citet{2020A&A...643A.148G} state that their \textit{Gaia} members in the Taurus, Upper Scorpius, and Lupus regions show a similar shape for the distribution of spectral types at the faint end of the initial mass function.

If the luminosity function is universal at least up to the Hyades age, then all our clusters must have been born with similar luminosity functions. However, the luminosity functions of our intermediate and old age clusters differ from the young clusters, and this might be due to many reasons. One reason is stellar evolution that, for example, causes massive stars to increase their luminosity when they turn into giants. On the other hand, since single stars of the same mass and metallicity evolve at the same rate, 
clusters of similar advanced age (e.g. the Hyades age) should still have similar luminosity functions. Our clusters are single (with the exception being the pair Collinder\,135 - UBC\,7, see Appendix \ref{sec.pair}), and the reason for the increased dissimilarity with age might be in their internal dynamical evolution, which depends on their density. However, further analysis with models and simulations would be needed to provide more evidence to support this claim.

\section{Conclusions} \label{sec.conclusions}
We selected 49 open clusters younger than 1\,Gyr and closer than 500\,pc and reassessed their membership using \textit{Gaia}\,DR3 and Hipparcos 6D astrometric information (positions, parallaxes, proper motions, and radial velocities) over the full range of validity of both missions. We discarded some known clusters in the Galactic plane. 
We provide the physical parameters of each cluster (3D position and velocity, distance, proper motion, age, tidal radius and mass, number of members) and an updated list of bona fide members, up to 1 tidal radius, and candidates, up to 3 tidal radii. We will make the new list of members public in a dedicated webpage\footnote{\url{http://research.iac.es/proyecto/gaiaclusters}} as well as on Vizier (the Centre de Donn\'ees de Strasbourg) for use by professional and amateur astronomers and educators. 
We studied the evolution of the luminosity function as a function of age and found that luminosity functions of the youngest clusters differ from the older populations, are more likely similar to each other, and show a greater degree of similarity than older clusters. We explained this observation with the universal luminosity function within the volume of our sample (500\,pc).
Clusters with ages similar to the Pleiades or Hyades show a lesser degree of similarity than younger clusters. If the luminosity function is universal, at least up to the age of the Hyades, one of the reasons for increased diversity of these single clusters could be internal dynamical evolution, but future work is needed to support this claim.

\begin{acknowledgements}
We thank the anonymous reviewer and acknowledge their valuable contribution that helped to significantly improve this paper.
M{\v Z} and NL acknowledge support from the Consejer{\'{\i}}a de Econom{\'{\i}}a, Conocimiento y Empleo del Gobierno de Canarias and the European Regional Development Fund (ERDF) under grant with reference PROID2020010052\@.
M{\v Z}, NL, JO, VB and ELM acknowledge support from the Agencia Estatal de Investigaci\'on del Ministerio de Ciencia e Innovaci\'on (AEI-MCINN) under grant PID2019-109522GB-C53\@.
APG acknowledges support from the grant PID2020-120052GB-I00 financed by MCIN/AEI/10.13039/501100011033\@.

Co-funded by the European Union (ERC, SUBSTELLAR, project number 101054354). Views and opinions expressed are however those of the author(s) only and do not necessarily reflect those of the European
Union or the European Research Council. Neither the European Union nor the granting authority can be held responsible for them.

This research has made use of the Simbad and Vizier databases, and the Aladin sky atlas  operated
at the centre de Donn\'ees Astronomiques de Strasbourg (CDS), and
of NASA's Astrophysics Data System Bibliographic Services (ADS). 
This research has made use of "Aladin sky atlas" developed at CDS, Strasbourg Observatory, France.
This work has made use of data from the European Space Agency (ESA) mission
{\it Gaia} (\url{https://www.cosmos.esa.int/gaia}), processed by the {\it Gaia}
Data Processing and Analysis Consortium (DPAC,
\url{https://www.cosmos.esa.int/web/gaia/dpac/consortium}). Funding for the DPAC
has been provided by national institutions, in particular the institutions
participating in the {\it Gaia} Multilateral Agreement.
This publication makes use of data products from the Two Micron All Sky Survey, which is a joint
project of the University of Massachusetts and the Infrared Processing and Analysis
Center/California Institute of Technology, funded by the National Aeronautics and Space
Administration and the National Science Foundation.\\
Based on observations obtained as part of the VISTA Hemisphere Survey, ESO Program, 179.A-2010 (PI: McMahon)
The UKIDSS project is defined in \citet{lawrence07}. UKIDSS uses the UKIRT Wide Field Camera
\citet[WFCAM;][]{2007A&A...467..777C}. The photometric system is described in \citet{2006MNRAS.367..454H}, and the
calibration is described in \citet{2009MNRAS.394..675H}. The pipeline processing and science archive are
described in Irwin et al.\ (2009, in prep) and \citet{2008MNRAS.384..637H}.\\
This publication makes use of data products from the Wide-field Infrared Survey Explorer, which
is a joint project of the University of California, Los Angeles, and the Jet Propulsion
Laboratory/California Institute of Technology, and NEOWISE, which is a project of the Jet
Propulsion Laboratory/California Institute of Technology. WISE and NEOWISE are funded
by the National Aeronautics and Space Administration. \\
The Pan-STARRS1 Surveys (PS1) and the PS1 public science archive have been made possible
through contributions by the Institute for Astronomy, the University of Hawaii, the Pan-STARRS
Project Office, the Max-Planck Society and its participating institutes, the Max Planck Institute
for Astronomy, Heidelberg and the Max Planck Institute for Extraterrestrial Physics, Garching,
The Johns Hopkins University, Durham University, the University of Edinburgh, the Queen's
University Belfast, the Harvard-Smithsonian Center for Astrophysics, the Las Cumbres Observatory
Global Telescope Network Incorporated, the National Central University of Taiwan, the Space
Telescope Science Institute, the National Aeronautics and Space Administration under
Grant No.\ NNX08AR22G issued through the Planetary Science Division of the NASA Science
Mission Directorate, the National Science Foundation Grant No. AST-1238877, the University
of Maryland, Eotvos Lorand University (ELTE), the Los Alamos National Laboratory, and the
Gordon and Betty Moore Foundation. \\
Funding for SDSS-III has been provided by the Alfred P. Sloan Foundation, the Participating Institutions, the National Science Foundation, and the U.S. Department of Energy Office of Science. The SDSS-III website is http://www.sdss3.org/.
SDSS-III is managed by the Astrophysical Research Consortium for the Participating Institutions of the SDSS-III Collaboration including the University of Arizona, the Brazilian Participation Group, Brookhaven National Laboratory, Carnegie Mellon University, University of Florida, the French Participation Group, the German Participation Group, Harvard University, the Instituto de Astrofisica de Canarias, the Michigan State/Notre Dame/JINA Participation Group, Johns Hopkins University, Lawrence Berkeley National Laboratory, Max Planck Institute for Astrophysics, Max Planck Institute for Extraterrestrial Physics, New Mexico State University, New York University, Ohio State University, Pennsylvania State University, University of Portsmouth, Princeton University, the Spanish Participation Group, University of Tokyo, University of Utah, Vanderbilt University, University of Virginia, University of Washington, and Yale University.

Software: \texttt{astropy} \citep{astropy:2018}, \texttt{NumPy} \citep{harris2020array}, \texttt{IPython} \citep{doi:10.1109/MCSE.2007.53}, \texttt{topcat} \citep{2005ASPC..347...29T} and \texttt{Matplotlib} \citep{doi:10.1109/MCSE.2007.55}. 
\end{acknowledgements}

%
%



\bibliographystyle{aa}
\bibliography{paper} 

\begin{appendix}

\section{Extinction} \label{sec.extinction}
\FloatBarrier
In Table \ref{tab.extinction}, we present model extinction and reddening values  for each bound cluster and compare them with the values in the literature. We estimated the upper E(B-V) limit for each star using the 2D maps of dust reddening with the \textit{GNILC} method from the \citealp{2016A&A...596A.109P} that is integrated in \texttt{dustmaps} \citep{2018JOSS....3..695M}. To avoid the model uncertainties for very hot and cool stars, we applied a photometric filter and only kept stars with Gaia's $0<B_P-R_P<1$. We assumed $R_\mathrm{V}=3.1$.

\onecolumn
\begin{table*}[htp]
\caption{Extinction and E(B-V) from \texttt{dustmaps} for the bound clusters.}
\begin{tabular}{l r | c c r | c l}
\hline
Cluster & Distance & E(B-V) & $A_\mathrm{V}$ & N & E(B-V) ref & Ref \\
\hline
 & pc & mag & mag &  & mag & \\
 \hline
Blanco 1        & 243 & 0.01 & 0.03 &  53 & 0.02 & \citet{westerlund88} \\
Melotte 111     &  86 & 0.01 & 0.04 &  29 & negligible & \citet{1972PASP...84..497F} \\
Alessi 13       & 103 & 0.02 & 0.06 &  12 & 0.02 & \citet{2022ApJS..262....7H}, derived from $A_\mathrm{0} = 0.05$ \\
Mamajek 4       & 441 & 0.07 & 0.21 &  24 &  & \\ 
Alessi 24       & 478 & 0.08 & 0.24 &  12 & 0.1 & \citet{2022ApJS..262....7H}, derived from $A_\mathrm{0} = 0.3$ \\
NGC 1039        & 492 & 0.09 & 0.27 &  71 & 0.12 & \citet{2023AJ....165...45M}, derived from $A_\mathrm{V} = 0.328$ \\
UBC 9           & 344 & 0.10 & 0.32 &  13 &  & \\ 
Alessi 10       & 434 & 0.14 & 0.43 &  14 & 0.2 &  \citet{2022ApJS..262....7H}, derived from $A_\mathrm{0} = 0.6$ \\
Alessi 3        & 275 & 0.14 & 0.44 &  10 & 0.16 & \citet{2022ApJS..262....7H}, derived from $A_\mathrm{0} = 0.5$ \\
NGC 2516        & 407 & 0.20 & 0.61 & 418 & 0.14 & \citet{1987MNRAS.228..483S} \\
Melotte 22      & 135 & 0.20 & 0.61 & 142 & 0.03 & \citet{1994MNRAS.268..181O} \\
UBC 7           & 274 & 0.21 & 0.64 &  11 &  & \\ 
NGC 1901        & 415 & 0.21 & 0.66 &  18 & 0.04 & \citet{carraro2007} \\
Melotte 25      &  46 & 0.48$^a$ & 1.50 &  82 & $\le 0.001$ & \citet{2006AJ....132.2453T} \\
IC 2391         & 151 & 0.58$^a$ & 1.79 &  34 & 0.006 & \citet{1996ApJS..106..489P} \\
Melotte 20      & 173 & 0.58$^a$ & 1.80 &  79 & negligible & \citet{2017ApJ...837...14A} \\
IC 2602         & 150 & 0.67$^a$ & 2.07 &  35 & 0 & \citet{arenou18}, most likely $A_\mathrm{g} = 0$ \\
\hline
\end{tabular}
\tablefoot{Some reference reddenings are determined from the reported extinction. In both cases, we assumed $R_\mathrm{V}=3.1$. \newline $^a$ These are upper limits. Most of the extinction comes from behind the clusters. This is true for all clusters in our list, but we emphasise this specifically for clusters with higher E(B-V) from \texttt{dustmaps}.
}
\label{tab.extinction}
\end{table*}
\twocolumn
\FloatBarrier

\section{Cluster properties}
Table \ref{tab.clusters_characteristics} presents the characteristics of bound clusters, while unbound objects are listed in Table \ref{tab.clusters_characteristics_unbound}. Cartesian coordinates and velocity dispersions are given in Table \ref{tab.clusters_characteristics_cartesian}.

\newcommand{\clustersstats}{IC 2602 & 160.7801 & $-$64.3706 & 289.6021 & $-$4.8724 & 150.39 $\pm$ 0.72 & 32.3 & 0.1 & 0.9 & 23.0 & 7.2 & 141 & 308 & 1 \\
Melotte 22 & 56.6517 & 24.0765 & 166.5277 & $-$23.6113 & 135.03 $\pm$ 0.26 & $-$16.2 & $-$2.6 & $-$30.5 & 32.2 & 11.2 & 529 & 1355 & 2 \\
Melotte 25 & 66.9682 & 16.5421 & 179.6565 & $-$21.7361 & 46.47 $\pm$ 0.14 & 140.8 & 39.0 & $-$33.1 & 46.1 & 8.4 & 217 & 427 & 3 \\
\hline
Blanco 1 & 0.8825 & $-$29.9150 & 15.2846 & $-$79.1200 & 234.34 $\pm$ 0.78 & $-$5.7 & $-$15.1 & $-$11.8 & 21.5 & 7.7 & 172 & 401 & 2 \\
NGC 1039 & 40.5334 & 42.7536 & 143.6702 & $-$15.6167 & 491.63 $\pm$ 1.90 & $-$5.4 & $-$1.5 & $-$9.2 & 16.0 & 7.8 & 176 & 217 & 2 \\
Alessi 13 & 51.4891 & $-$35.8355 & 237.5839 & $-$56.1610 & 103.46 $\pm$ 0.88 & $-$7.6 & $-$37.7 & $-$17.4 & 25.7 & 10.0 & 27 & 71 & 1 \\
Melotte 20 & 51.5358 & 48.9839 & 147.3072 & $-$6.4265 & 173.27 $\pm$ 0.53 & $-$6.4 & 19.5 & $-$22.7 & 28.3 & 8.8 & 252 & 524 & 2 \\
Platais 3 & 68.8478 & 71.2787 & 138.9824 & 15.7877 & 177.99 $\pm$ 1.87 & 17.8 & 9.9 & $-$9.2 & 19.5 & 4.0 & 24 & 38 & 2 \\
NGC 1901 & 79.8301 & $-$68.3214 & 278.8643 & $-$33.5280 & 415.22 $\pm$ 4.15 & 12.0 & $-$4.3 & $-$1.5 & 25.2 & 4.2 & 28 & 23 & 3 \\
UBC 7 & 106.6258 & $-$37.6535 & 248.6182 & $-$13.4307 & 273.92 $\pm$ 1.74 & 14.2 & $-$7.4 & 5.4 & 21.4 & 4.6 & 37 & 92 & 1 \\
Alessi 3 & 109.2165 & $-$46.3786 & 257.6573 & $-$15.1952 & 275.25 $\pm$ 2.65 & 12.2 & 3.3 & 11.6 & 20.2 & 3.8 & 20 & 24 & 3 \\
Collinder 135 & 109.3361 & $-$36.7609 & 248.7192 & $-$11.0970 & 293.88 $\pm$ 1.77 & 13.6 & $-$6.9 & 6.6 & 22.2 & 5.4 & 57 & 91 & 1 \\
ASCC 41 & 116.7218 & 0.0704 & 219.2884 & 12.3447 & 293.08 $\pm$ 2.55 & $-$7.7 & $-$2.8 & $-$3.1 & 11.8 & 4.5 & 35 & 44 & 2 \\
NGC 2516 & 119.5561 & $-$60.7628 & 273.8338 & $-$15.8432 & 407.27 $\pm$ 0.55 & 15.9 & $-$5.6 & $-$1.3 & 32.6 & 14.6 & 1152 & 1894 & 2 \\
NGC 2632 & 129.9961 & 19.6375 & 205.9118 & 32.3824 & 183.35 $\pm$ 0.42 & 52.7 & $-$14.6 & 8.0 & 48.2 & 10.6 & 442 & 925 & 3 \\
IC 2391 & 130.2486 & $-$53.0266 & 270.3989 & $-$6.7794 & 150.54 $\pm$ 0.53 & 37.4 & 1.2 & 7.5 & 27.7 & 6.9 & 124 & 270 & 1 \\
Melotte 111 & 185.8656 & 26.0623 & 221.0263 & 83.6569 & 85.53 $\pm$ 0.55 & 12.9 & $-$6.0 & 2.1 & 6.1 & 5.2 & 52 & 74 & 3 \\
Alessi 24 & 260.8358 & $-$62.7086 & 328.9885 & $-$14.6079 & 477.81 $\pm$ 4.78 & $-$0.8 & 5.0 & 18.4 & 21.7 & 5.0 & 31 & 31 & 2 \\
Mamajek 4 & 276.5463 & $-$50.9501 & 343.8369 & $-$17.0053 & 441.17 $\pm$ 3.10 & $-$22.9 & 2.8 & 25.1 & 54.3 & 10.0 & 53 & 61 & 3 \\
UBC 9 & 276.7157 & 26.4682 & 54.5648 & 16.7681 & 343.69 $\pm$ 3.19 & $-$3.7 & $-$10.3 & 11.8 & 21.5 & 7.5 & 28 & 47 & 1 \\
Stephenson 1 & 283.5958 & 36.8612 & 66.8801 & 15.3303 & 354.55 $\pm$ 1.97 & $-$3.9 & $-$9.5 & 7.4 & 18.8 & 5.8 & 73 & 123 & 1 \\
ASCC 101 & 288.3179 & 36.3917 & 68.0206 & 11.6767 & 392.43 $\pm$ 3.58 & $-$7.4 & $-$5.6 & 7.0 & 18.7 & 4.4 & 31 & 36 & 2 \\
Alessi 10 & 301.2260 & $-$10.4884 & 31.6592 & $-$21.0234 & 433.65 $\pm$ 3.63 & $-$9.3 & $-$0.2 & 13.9 & 23.7 & 10.0 & 40 & 58 & 2 \\}
\begin{landscape}
\begin{table}
\caption{Summary of the cluster properties for the bound subsample.}
\begin{tabular}{l rr rr r | rr rr | r rr r}
\hline
Cluster & $\alpha$ & $\delta$ & l & b & d & $\mu_\alpha*$ & $\mu_\delta$ & RV & v &  $r_\mathrm{t}$ & M(tidal) & N & Age group \\
\hline
 & deg & deg & deg & deg & pc & mas\,yr$^{-1}$ & mas\,yr$^{-1}$ & km\,s$^{-1}$ & km\,s$^{-1}$ & pc & M$_\odot$ & & \\
 \hline
 IC 2602 & 160.7801 & $-$64.3706 & 289.6021 & $-$4.8724 & 150.39 $\pm$ 0.72 & 32.3 & 0.1 & 0.9 & 23.0 & 7.2 & 141 & 308 & 1 \\
Melotte 22 & 56.6517 & 24.0765 & 166.5277 & $-$23.6113 & 135.03 $\pm$ 0.26 & $-$16.2 & $-$2.6 & $-$30.5 & 32.2 & 11.2 & 529 & 1355 & 2 \\
Melotte 25 & 66.9682 & 16.5421 & 179.6565 & $-$21.7361 & 46.47 $\pm$ 0.14 & 140.8 & 39.0 & $-$33.1 & 46.1 & 8.4 & 217 & 427 & 3 \\
\hline
Blanco 1 & 0.8825 & $-$29.9150 & 15.2846 & $-$79.1200 & 234.34 $\pm$ 0.78 & $-$5.7 & $-$15.1 & $-$11.8 & 21.5 & 7.7 & 172 & 401 & 2 \\
NGC 1039 & 40.5334 & 42.7536 & 143.6702 & $-$15.6167 & 491.63 $\pm$ 1.90 & $-$5.4 & $-$1.5 & $-$9.2 & 16.0 & 7.8 & 176 & 217 & 2 \\
Alessi 13 & 51.4891 & $-$35.8355 & 237.5839 & $-$56.1610 & 103.46 $\pm$ 0.88 & $-$7.6 & $-$37.7 & $-$17.4 & 25.7 & 10.0 & 27 & 71 & 1 \\
Melotte 20 & 51.5358 & 48.9839 & 147.3072 & $-$6.4265 & 173.27 $\pm$ 0.53 & $-$6.4 & 19.5 & $-$22.7 & 28.3 & 8.8 & 252 & 524 & 2 \\
Platais 3 & 68.8478 & 71.2787 & 138.9824 & 15.7877 & 177.99 $\pm$ 1.87 & 17.8 & 9.9 & $-$9.2 & 19.5 & 4.0 & 24 & 38 & 2 \\
NGC 1901 & 79.8301 & $-$68.3214 & 278.8643 & $-$33.5280 & 415.22 $\pm$ 4.15 & 12.0 & $-$4.3 & $-$1.5 & 25.2 & 4.2 & 28 & 23 & 3 \\
UBC 7 & 106.6258 & $-$37.6535 & 248.6182 & $-$13.4307 & 273.92 $\pm$ 1.74 & 14.2 & $-$7.4 & 5.4 & 21.4 & 4.6 & 37 & 92 & 1 \\
Alessi 3 & 109.2165 & $-$46.3786 & 257.6573 & $-$15.1952 & 275.25 $\pm$ 2.65 & 12.2 & 3.3 & 11.6 & 20.2 & 3.8 & 20 & 24 & 3 \\
Collinder 135 & 109.3361 & $-$36.7609 & 248.7192 & $-$11.0970 & 293.88 $\pm$ 1.77 & 13.6 & $-$6.9 & 6.6 & 22.2 & 5.4 & 57 & 91 & 1 \\
ASCC 41 & 116.7218 & 0.0704 & 219.2884 & 12.3447 & 293.08 $\pm$ 2.55 & $-$7.7 & $-$2.8 & $-$3.1 & 11.8 & 4.5 & 35 & 44 & 2 \\
NGC 2516 & 119.5561 & $-$60.7628 & 273.8338 & $-$15.8432 & 407.27 $\pm$ 0.55 & 15.9 & $-$5.6 & $-$1.3 & 32.6 & 14.6 & 1152 & 1894 & 2 \\
NGC 2632 & 129.9961 & 19.6375 & 205.9118 & 32.3824 & 183.35 $\pm$ 0.42 & 52.7 & $-$14.6 & 8.0 & 48.2 & 10.6 & 442 & 925 & 3 \\
IC 2391 & 130.2486 & $-$53.0266 & 270.3989 & $-$6.7794 & 150.54 $\pm$ 0.53 & 37.4 & 1.2 & 7.5 & 27.7 & 6.9 & 124 & 270 & 1 \\
Melotte 111 & 185.8656 & 26.0623 & 221.0263 & 83.6569 & 85.53 $\pm$ 0.55 & 12.9 & $-$6.0 & 2.1 & 6.1 & 5.2 & 52 & 74 & 3 \\
Alessi 24 & 260.8358 & $-$62.7086 & 328.9885 & $-$14.6079 & 477.81 $\pm$ 4.78 & $-$0.8 & 5.0 & 18.4 & 21.7 & 5.0 & 31 & 31 & 2 \\
Mamajek 4 & 276.5463 & $-$50.9501 & 343.8369 & $-$17.0053 & 441.17 $\pm$ 3.10 & $-$22.9 & 2.8 & 25.1 & 54.3 & 10.0 & 53 & 61 & 3 \\
UBC 9 & 276.7157 & 26.4682 & 54.5648 & 16.7681 & 343.69 $\pm$ 3.19 & $-$3.7 & $-$10.3 & 11.8 & 21.5 & 7.5 & 28 & 47 & 1 \\
Stephenson 1 & 283.5958 & 36.8612 & 66.8801 & 15.3303 & 354.55 $\pm$ 1.97 & $-$3.9 & $-$9.5 & 7.4 & 18.8 & 5.8 & 73 & 123 & 1 \\
ASCC 101 & 288.3179 & 36.3917 & 68.0206 & 11.6767 & 392.43 $\pm$ 3.58 & $-$7.4 & $-$5.6 & 7.0 & 18.7 & 4.4 & 31 & 36 & 2 \\
Alessi 10 & 301.2260 & $-$10.4884 & 31.6592 & $-$21.0234 & 433.65 $\pm$ 3.63 & $-$9.3 & $-$0.2 & 13.9 & 23.7 & 10.0 & 40 & 58 & 2 \\
 \hline
\end{tabular}
\tablefoot{Benchmark clusters are listed on top. We provide distance (d), cluster proper motions $\mu_\alpha*$ and $\mu_\delta$, line-of-sight velocity (RV), systemic velocity (v), tidal radius $r_\mathrm{t}$, tidal mass M(tidal), and the number of bona fide members (\texttt{bona\_fide\_members}). Age groups are denoted with 1 (age of IC\,2602), 2 (age of Melotte\,22\,=\,Pleiades), and 3 (age of Melotte\,25\,=\,Hyades).
}
\label{tab.clusters_characteristics}
\end{table}
\end{landscape}

\newcommand{\clustersstatsunbound}{UPK 303 & 39.3939 & 40.6252 & 143.8065 & $-$17.9051 & 210.99 $\pm$ 2.22 & $-$18.8 & 6.8 & $-$24.3 & 31.5 & 10.0 & 24 & 49 \\
UPK 305 & 40.3849 & 38.8866 & 145.3155 & $-$19.1525 & 400.13 $\pm$ 10.18 & $-$3.7 & $-$0.3 & $-$11.9 & 13.8 & 10.0 & 15 & 26 \\
UBC 31 & 61.0929 & 32.8044 & 163.2998 & $-$14.5508 & 356.21 $\pm$ 1.63 & 7.8 & 1.2 & $-$15.3 & 20.3 & 20.0 & 156 & 500 \\
NGC 1662 & 72.1526 & 10.8962 & 187.7547 & $-$21.1051 & 402.64 $\pm$ 2.62 & $-$7.0 & 0.2 & 4.6 & 14.1 & 5.1 & 49 & 67 \\
UPK 385 & 75.5438 & 12.9241 & 187.9761 & $-$17.2222 & 313.00 $\pm$ 5.70 & 7.0 & $-$1.5 & $-$11.4 & 15.6 & 5.0 & 7 & 15 \\
Platais 4 & 77.1983 & 22.3997 & 180.8842 & $-$10.5194 & 313.50 $\pm$ 26.86 &   &   &   &   & 20.0 & 64 & 216 \\
Mamajek 3 & 81.8939 & 6.6251 & 197.0021 & $-$15.2903 & 98.46 $\pm$ 2.32 & 18.1 & $-$13.0 & $-$21.9 & 24.2 & 2.6 & 7 & 7 \\
UBC 8 & 84.4349 & 57.1315 & 154.8780 & 13.3166 & 471.86 $\pm$ 2.70 & $-$1.4 & $-$0.9 & $-$13.2 & 13.7 & 10.0 & 69 & 103 \\
UPK 350 & 101.7852 & 48.6801 & 167.3346 & 19.3564 & 426.83 $\pm$ 4.40 & 4.1 & $-$2.0 & $-$13.3 & 16.1 & 10.0 & 20 & 28 \\
UPK 540 & 114.8198 & $-$58.6372 & 270.7512 & $-$17.0598 & 362.16 $\pm$ 3.11 & 9.6 & $-$0.5 & 4.1 & 16.9 & 10.0 & 29 & 54 \\
UPK 533 & 123.2414 & $-$52.4858 & 267.5120 & $-$9.9585 & 336.57 $\pm$ 3.00 & 13.3 & $-$10.8 & 1.0 & 27.4 & 10.0 & 19 & 44 \\
UBC 12 & 126.0640 & $-$8.5525 & 231.7665 & 16.1980 & 437.58 $\pm$ 9.40 & 17.4 & $-$0.0 & 7.1 & 36.8 & 15.0 & 23 & 29 \\
Mamajek 1 & 130.6553 & $-$78.3476 & 291.8782 & $-$21.2700 & 100.91 $\pm$ 1.78 & 44.2 & $-$16.7 & 9.0 & 24.3 & 50.0 & 14 & 35 \\
Turner 5 & 142.9664 & $-$36.4349 & 264.1163 & 10.9797 & 416.73 $\pm$ 3.87 & $-$2.1 & $-$2.6 & 1.1 & 6.7 & 10.0 & 21 & 25 \\
UPK 606 & 215.7553 & $-$46.4112 & 318.9521 & 13.5922 & 168.28 $\pm$ 2.09 & 15.4 & 11.6 & 15.8 & 22.1 & 3.0 & 10 & 26 \\
UPK 612 & 254.1704 & $-$68.3064 & 322.5872 & $-$15.4121 & 225.14 $\pm$ 1.47 & $-$3.9 & 17.6 & 4.0 & 19.7 & 15.0 & 89 & 200 \\
UPK 624 & 254.3194 & $-$60.5082 & 329.0544 & $-$10.8109 & 311.80 $\pm$ 4.02 & $-$3.0 & 6.7 & 18.6 & 21.5 & 10.0 & 24 & 45 \\
Mamajek 2 & 264.3584 & $-$7.9914 & 17.0597 & 12.4905 & 195.07 $\pm$ 2.32 & $-$8.7 & $-$1.2 & 27.4 & 28.6 & 5.0 & 19 & 40 \\
IC 4665 & 266.3973 & 5.6721 & 30.4945 & 17.2198 & 340.89 $\pm$ 3.13 & $-$1.2 & $-$6.3 & 16.4 & 19.4 & 4.5 & 34 & 50 \\
Collinder 350 & 267.0184 & 1.4074 & 26.8441 & 14.7183 & 366.00 $\pm$ 3.28 & $-$7.2 & 1.7 & 11.8 & 17.4 & 5.0 & 27 & 34 \\
Gulliver 20 & 273.6913 & 11.1062 & 38.8340 & 13.1195 & 419.82 $\pm$ 5.02 & $-$2.6 & $-$7.0 & 14.5 & 20.8 & 10.0 & 33 & 51 \\
UPK 654 & 277.2507 & $-$34.5111 & 359.5211 & $-$10.7585 & 495.99 $\pm$ 6.28 & $-$1.0 & 0.0 & 20.7 & 20.8 & 10.0 & 37 & 58 \\
UBC 1 & 287.9726 & 56.8465 & 87.5620 & 19.7780 & 323.18 $\pm$ 1.78 & $-$7.9 & $-$12.6 & 9.1 & 24.5 & 10.0 & 67 & 115 \\
UPK 88 & 309.8308 & 21.1573 & 64.6697 & $-$12.2315 & 285.23 $\pm$ 4.12 & $-$4.4 & $-$10.2 & $-$5.1 & 15.9 & 7.5 & 9 & 15 \\
UBC 159 & 339.0372 & 39.6582 & 96.4360 & $-$16.1532 & 489.72 $\pm$ 2.71 & $-$5.0 & $-$5.4 & 11.6 & 20.6 & 15.0 & 76 & 183 \\
Aveni$-$Hunter 1 & 354.3071 & 48.3612 & 110.4164 & $-$12.6989 & 416.44 $\pm$ 5.62 & $-$7.4 & $-$3.1 & 4.9 & 16.6 & 5.0 & 14 & 24 \\}
\begin{landscape}
\begin{table}
\caption{Summary of the cluster properties for clusters in the unbound subsample.}
\begin{tabular}{l rr rr r | rr rr | r rr}
\hline
Cluster & $\alpha$ & $\delta$ & l & b & d & $\mu_\alpha*$ & $\mu_\delta$ & RV & v &  $r_\mathrm{t}$ & M(tidal) & N \\
\hline
 & deg & deg & deg & deg & pc & mas\,yr$^{-1}$ & mas\,yr$^{-1}$ & km\,s$^{-1}$ & km\,s$^{-1}$ & pc & M$_\odot$ & \\
 \hline
 UPK 303 & 39.3939 & 40.6252 & 143.8065 & $-$17.9051 & 210.99 $\pm$ 2.22 & $-$18.8 & 6.8 & $-$24.3 & 31.5 & 10.0 & 24 & 49 \\
UPK 305 & 40.3849 & 38.8866 & 145.3155 & $-$19.1525 & 400.13 $\pm$ 10.18 & $-$3.7 & $-$0.3 & $-$11.9 & 13.8 & 10.0 & 15 & 26 \\
UBC 31 & 61.0929 & 32.8044 & 163.2998 & $-$14.5508 & 356.21 $\pm$ 1.63 & 7.8 & 1.2 & $-$15.3 & 20.3 & 20.0 & 156 & 500 \\
NGC 1662 & 72.1526 & 10.8962 & 187.7547 & $-$21.1051 & 402.64 $\pm$ 2.62 & $-$7.0 & 0.2 & 4.6 & 14.1 & 5.1 & 49 & 67 \\
UPK 385 & 75.5438 & 12.9241 & 187.9761 & $-$17.2222 & 313.00 $\pm$ 5.70 & 7.0 & $-$1.5 & $-$11.4 & 15.6 & 5.0 & 7 & 15 \\
Platais 4 & 77.1983 & 22.3997 & 180.8842 & $-$10.5194 & 313.50 $\pm$ 26.86 &   &   &   &   & 20.0 & 64 & 216 \\
Mamajek 3 & 81.8939 & 6.6251 & 197.0021 & $-$15.2903 & 98.46 $\pm$ 2.32 & 18.1 & $-$13.0 & $-$21.9 & 24.2 & 2.6 & 7 & 7 \\
UBC 8 & 84.4349 & 57.1315 & 154.8780 & 13.3166 & 471.86 $\pm$ 2.70 & $-$1.4 & $-$0.9 & $-$13.2 & 13.7 & 10.0 & 69 & 103 \\
UPK 350 & 101.7852 & 48.6801 & 167.3346 & 19.3564 & 426.83 $\pm$ 4.40 & 4.1 & $-$2.0 & $-$13.3 & 16.1 & 10.0 & 20 & 28 \\
UPK 540 & 114.8198 & $-$58.6372 & 270.7512 & $-$17.0598 & 362.16 $\pm$ 3.11 & 9.6 & $-$0.5 & 4.1 & 16.9 & 10.0 & 29 & 54 \\
UPK 533 & 123.2414 & $-$52.4858 & 267.5120 & $-$9.9585 & 336.57 $\pm$ 3.00 & 13.3 & $-$10.8 & 1.0 & 27.4 & 10.0 & 19 & 44 \\
UBC 12 & 126.0640 & $-$8.5525 & 231.7665 & 16.1980 & 437.58 $\pm$ 9.40 & 17.4 & $-$0.0 & 7.1 & 36.8 & 15.0 & 23 & 29 \\
Mamajek 1 & 130.6553 & $-$78.3476 & 291.8782 & $-$21.2700 & 100.91 $\pm$ 1.78 & 44.2 & $-$16.7 & 9.0 & 24.3 & 50.0 & 14 & 35 \\
Turner 5 & 142.9664 & $-$36.4349 & 264.1163 & 10.9797 & 416.73 $\pm$ 3.87 & $-$2.1 & $-$2.6 & 1.1 & 6.7 & 10.0 & 21 & 25 \\
UPK 606 & 215.7553 & $-$46.4112 & 318.9521 & 13.5922 & 168.28 $\pm$ 2.09 & 15.4 & 11.6 & 15.8 & 22.1 & 3.0 & 10 & 26 \\
UPK 612 & 254.1704 & $-$68.3064 & 322.5872 & $-$15.4121 & 225.14 $\pm$ 1.47 & $-$3.9 & 17.6 & 4.0 & 19.7 & 15.0 & 89 & 200 \\
UPK 624 & 254.3194 & $-$60.5082 & 329.0544 & $-$10.8109 & 311.80 $\pm$ 4.02 & $-$3.0 & 6.7 & 18.6 & 21.5 & 10.0 & 24 & 45 \\
Mamajek 2 & 264.3584 & $-$7.9914 & 17.0597 & 12.4905 & 195.07 $\pm$ 2.32 & $-$8.7 & $-$1.2 & 27.4 & 28.6 & 5.0 & 19 & 40 \\
IC 4665 & 266.3973 & 5.6721 & 30.4945 & 17.2198 & 340.89 $\pm$ 3.13 & $-$1.2 & $-$6.3 & 16.4 & 19.4 & 4.5 & 34 & 50 \\
Collinder 350 & 267.0184 & 1.4074 & 26.8441 & 14.7183 & 366.00 $\pm$ 3.28 & $-$7.2 & 1.7 & 11.8 & 17.4 & 5.0 & 27 & 34 \\
Gulliver 20 & 273.6913 & 11.1062 & 38.8340 & 13.1195 & 419.82 $\pm$ 5.02 & $-$2.6 & $-$7.0 & 14.5 & 20.8 & 10.0 & 33 & 51 \\
UPK 654 & 277.2507 & $-$34.5111 & 359.5211 & $-$10.7585 & 495.99 $\pm$ 6.28 & $-$1.0 & 0.0 & 20.7 & 20.8 & 10.0 & 37 & 58 \\
UBC 1 & 287.9726 & 56.8465 & 87.5620 & 19.7780 & 323.18 $\pm$ 1.78 & $-$7.9 & $-$12.6 & 9.1 & 24.5 & 10.0 & 67 & 115 \\
UPK 88 & 309.8308 & 21.1573 & 64.6697 & $-$12.2315 & 285.23 $\pm$ 4.12 & $-$4.4 & $-$10.2 & $-$5.1 & 15.9 & 7.5 & 9 & 15 \\
UBC 159 & 339.0372 & 39.6582 & 96.4360 & $-$16.1532 & 489.72 $\pm$ 2.71 & $-$5.0 & $-$5.4 & 11.6 & 20.6 & 15.0 & 76 & 183 \\
Aveni$-$Hunter 1 & 354.3071 & 48.3612 & 110.4164 & $-$12.6989 & 416.44 $\pm$ 5.62 & $-$7.4 & $-$3.1 & 4.9 & 16.6 & 5.0 & 14 & 24 \\
 \hline
\end{tabular}
\tablefoot{We provide distance (d), cluster proper motions $\mu_\alpha*$ and $\mu_\delta$, line-of-sight velocity (RV), systemic velocity (v), tidal radius $r_\mathrm{t}$, tidal mass M(tidal), and the number of bona fide members (\texttt{bona\_fide\_members}).
}
\label{tab.clusters_characteristics_unbound}
\end{table}
\end{landscape}

\onecolumn
\newcommand{\clustersstatscart}{Blanco 1 & 42.67 & 0.21 & 11.66 & 0.45 & $-$230.13 & 0.73 & $-$18.53 & 0.19 & $-$6.61 & 0.06 & $-$8.67 & 0.98 & 0.83 & 0.54 & 0.63 & 10.59 \\
UPK 303 & $-$162.03 & 1.28 & 118.56 & 1.43 & $-$64.87 & 1.39 & $-$5.75 & 3.52 & $-$29.11 & 2.68 & $-$10.68 & 1.41 & 1.26 & 0.67 & 1.06 & 15.29 \\
UPK 305 & $-$310.81 & 7.36 & 215.09 & 5.73 & $-$131.28 & 4.21 & $-$2.27 & 1.31 & $-$11.10 & 0.93 & $-$7.88 & 0.59 & 4.10 & 0.89 & 4.00 & 4.16 \\
NGC 1039 & $-$381.45 & 1.16 & 280.51 & 1.23 & $-$132.35 & 1.10 & 4.84 & 0.69 & $-$12.44 & 0.52 & $-$8.84 & 0.24 & 1.14 & 0.70 & 0.90 & 9.40 \\
Alessi 13 & $-$30.88 & 0.60 & $-$48.64 & 0.56 & $-$85.93 & 0.52 & $-$12.61 & 0.49 & $-$21.82 & 0.73 & $-$4.82 & 1.32 & 1.10 & 0.79 & 0.77 & 8.93 \\
Melotte 20 & $-$144.90 & 0.36 & 93.00 & 0.37 & $-$19.39 & 0.28 & $-$12.71 & 0.91 & $-$24.39 & 0.58 & $-$6.58 & 0.13 & 3.66 & 1.20 & 3.46 & 14.78 \\
Melotte 22 & $-$120.32 & 0.19 & 28.82 & 0.14 & $-$54.08 & 0.13 & $-$6.28 & 0.43 & $-$28.36 & 0.11 & $-$13.93 & 0.20 & 1.75 & 0.76 & 1.57 & 10.70 \\
UBC 31 & $-$330.25 & 1.37 & 99.08 & 0.73 & $-$89.49 & 0.62 & $-$18.25 & 1.63 & $-$5.90 & 0.49 & $-$6.61 & 0.45 & 7.04 & 2.93 & 6.40 & 16.31 \\
Melotte 25 & $-$43.16 & 0.12 & 0.26 & 0.07 & $-$17.21 & 0.06 & $-$41.91 & 0.34 & $-$19.28 & 0.06 & $-$1.18 & 0.17 & 3.67 & 2.57 & 2.62 & 6.59 \\
Platais 3 & $-$129.23 & 1.03 & 112.41 & 1.40 & 48.43 & 0.86 & $-$17.95 & 0.69 & $-$4.75 & 0.57 & $-$6.03 & 0.22 & 0.70 & 0.59 & 0.38 & 3.65 \\
NGC 1662 & $-$372.20 & 2.30 & $-$50.69 & 1.24 & $-$144.99 & 0.93 & 14.01 & 0.79 & 0.12 & 0.12 & 1.16 & 0.30 & 0.66 & 0.47 & 0.46 & 5.25 \\
UPK 385 & $-$296.07 & 5.10 & $-$41.48 & 2.57 & $-$92.67 & 1.80 & $-$12.67 & 2.70 & $-$7.69 & 0.40 & $-$4.76 & 0.88 & 1.71 & 0.62 & 1.59 & 6.41 \\
Platais 4 & $-$308.19 & 26.23 & $-$4.76 & 4.72 & $-$57.23 & 5.53 &   &   &   &   &   &   & 6.50 & 2.07 & 6.16 & 10.49 \\
NGC 1901 & 53.34 & 2.51 & $-$342.00 & 3.29 & $-$229.34 & 2.01 & $-$24.77 & 0.07 & $-$4.03 & 0.58 & $-$1.72 & 0.39 & 0.47 & 0.37 & 0.28 & 2.77 \\
Mamajek 3 & $-$90.82 & 2.03 & $-$27.77 & 1.07 & $-$25.96 & 0.56 & $-$11.34 & 0.16 & $-$19.63 & 0.43 & $-$8.55 & 0.03 & 0.29 & 0.14 & 0.25 & 0.50 \\
UBC 8 & $-$415.74 & 1.94 & 194.94 & 1.81 & 108.69 & 1.07 & 2.63 & 0.75 & $-$5.84 & 0.35 & $-$12.16 & 0.21 & 1.35 & 0.96 & 0.95 & 6.19 \\
UPK 350 & $-$392.91 & 3.44 & 88.30 & 2.98 & 141.47 & 1.86 & $-$6.92 & 0.72 & $-$7.34 & 0.18 & $-$12.59 & 0.28 & 0.37 & 0.23 & 0.29 & 2.67 \\
UBC 7 & $-$97.13 & 1.43 & $-$248.09 & 1.17 & $-$63.62 & 0.73 & $-$17.16 & 0.88 & $-$6.81 & 2.20 & $-$10.86 & 0.58 & 0.53 & 0.37 & 0.38 & 11.90 \\
Alessi 3 & $-$56.78 & 2.11 & $-$259.49 & 1.99 & $-$72.14 & 1.28 & $-$18.72 & 0.11 & 5.23 & 0.36 & $-$5.47 & 0.12 & 0.37 & 0.26 & 0.27 & 1.32 \\
Collinder 135 & $-$104.67 & 1.45 & $-$268.72 & 1.21 & $-$56.56 & 0.78 & $-$17.66 & 0.59 & $-$6.69 & 1.42 & $-$11.59 & 0.36 & 0.86 & 0.65 & 0.56 & 9.56 \\
UPK 540 & 4.54 & 2.06 & $-$346.19 & 2.73 & $-$106.25 & 1.61 & $-$15.50 & 0.03 & $-$5.56 & 2.54 & $-$3.89 & 0.79 & 3.66 & 1.80 & 3.19 & 10.22 \\
ASCC 41 & $-$221.59 & 2.06 & $-$181.30 & 1.22 & 62.66 & 1.12 & 10.95 & 0.91 & 2.03 & 0.75 & $-$3.91 & 0.26 & 0.44 & 0.30 & 0.31 & 6.16 \\
NGC 2516 & 26.20 & 0.32 & $-$390.93 & 0.47 & $-$111.19 & 0.27 & $-$21.81 & 0.04 & $-$23.93 & 0.25 & $-$4.10 & 0.08 & 2.44 & 1.43 & 1.97 & 6.46 \\
UPK 533 & $-$14.39 & 1.95 & $-$331.18 & 2.68 & $-$58.20 & 1.60 & $-$10.54 & 0.19 & $-$22.54 & 2.93 & $-$11.36 & 0.42 & 5.27 & 3.45 & 3.99 & 9.76 \\
UBC 12 & $-$260.05 & 5.96 & $-$330.07 & 6.49 & 122.07 & 3.44 & $-$33.37 & 0.58 & $-$15.59 & 0.77 & $-$1.15 & 0.25 & 2.94 & 2.52 & 1.50 & 3.50 \\
NGC 2632 & $-$139.27 & 0.28 & $-$67.66 & 0.21 & 98.20 & 0.24 & $-$42.69 & 0.20 & $-$20.39 & 0.10 & $-$9.28 & 0.14 & 1.64 & 1.20 & 1.12 & 5.17 \\
IC 2391 & 1.04 & 0.39 & $-$149.49 & 0.49 & $-$17.77 & 0.35 & $-$23.71 & 0.06 & $-$13.25 & 1.17 & $-$5.46 & 0.15 & 1.75 & 0.84 & 1.54 & 11.17 \\
Mamajek 1 & 35.04 & 0.79 & $-$87.26 & 1.41 & $-$36.61 & 0.78 & $-$12.13 & 0.27 & $-$18.31 & 1.07 & $-$10.46 & 0.39 & 3.47 & 2.18 & 2.69 & 4.15 \\
Turner 5 & $-$41.94 & 2.51 & $-$406.95 & 3.11 & 79.37 & 3.07 & 4.23 & 0.14 & 2.03 & 0.58 & $-$4.78 & 0.22 & 1.27 & 0.37 & 1.22 & 2.35 \\
IC 2602 & 50.27 & 0.33 & $-$141.16 & 0.62 & $-$12.77 & 0.35 & $-$8.02 & 0.50 & $-$21.56 & 1.39 & $-$0.81 & 0.14 & 2.11 & 0.65 & 2.01 & 15.59 \\
Melotte 111 & $-$7.13 & 0.15 & $-$6.20 & 0.35 & 85.01 & 0.52 & $-$2.41 & 0.06 & $-$5.50 & 0.06 & $-$1.28 & 0.65 & 0.43 & 0.32 & 0.29 & 4.50 \\
UPK 606 & 123.36 & 1.16 & $-$107.42 & 1.46 & 39.55 & 1.35 & $-$7.11 & 3.62 & $-$20.25 & 3.15 & $-$5.08 & 1.17 & 9.95 & 9.42 & 3.20 & 17.12 \\
UPK 612 & 172.40 & 0.97 & $-$131.87 & 0.98 & $-$59.83 & 0.56 & $-$9.14 & 0.62 & $-$17.09 & 0.45 & 3.25 & 0.23 & 2.86 & 1.75 & 2.27 & 5.91 \\
UPK 624 & 262.67 & 2.86 & $-$157.49 & 2.61 & $-$58.48 & 1.51 & $-$9.08 & 1.71 & $-$15.84 & 1.04 & $-$11.34 & 0.41 & 5.33 & 1.25 & 5.18 & 6.77 \\
Alessi 24 & 396.28 & 3.13 & $-$238.22 & 3.36 & $-$120.50 & 2.01 & $-$4.77 & 0.62 & $-$17.94 & 0.39 & $-$11.21 & 0.22 & 0.65 & 0.43 & 0.49 & 3.24 \\
Mamajek 2 & 182.07 & 2.04 & 55.87 & 1.05 & 42.19 & 0.52 & $-$10.64 & 1.85 & $-$26.10 & 0.57 & $-$4.88 & 0.46 & 1.14 & 0.68 & 0.91 & 7.51 \\
IC 4665 & 280.57 & 2.73 & 165.23 & 1.42 & 100.92 & 0.67 & $-$3.09 & 1.47 & $-$17.15 & 0.88 & $-$8.52 & 0.53 & 0.62 & 0.46 & 0.41 & 8.34 \\
Collinder 350 & 315.85 & 2.86 & 159.85 & 1.46 & 92.99 & 0.67 & $-$13.04 & 1.25 & $-$11.09 & 0.64 & 3.23 & 0.38 & 0.52 & 0.42 & 0.30 & 7.10 \\
Gulliver 20 & 318.49 & 4.05 & 256.38 & 2.77 & 95.29 & 1.11 & $-$4.15 & 1.18 & $-$17.23 & 0.93 & $-$10.93 & 0.33 & 3.06 & 1.14 & 2.84 & 6.86 \\
Mamajek 4 & 405.21 & 2.37 & $-$117.44 & 2.06 & $-$129.03 & 1.25 & $-$45.18 & 0.59 & $-$25.62 & 0.23 & $-$15.82 & 0.23 & 1.94 & 1.31 & 1.44 & 3.93 \\
UBC 9 & 190.79 & 2.69 & 268.12 & 1.82 & 99.16 & 0.96 & $-$3.92 & 1.08 & $-$18.69 & 1.50 & $-$9.79 & 0.58 & 1.89 & 0.42 & 1.85 & 8.64 \\
UPK 654 & 487.26 & 6.07 & $-$4.07 & 2.12 & $-$92.59 & 1.56 & $-$0.19 & 4.11 & $-$17.25 & 0.11 & $-$11.60 & 0.82 & 4.14 & 1.16 & 3.98 & 16.71 \\
Stephenson 1 & 134.26 & 1.60 & 314.47 & 1.31 & 93.74 & 0.77 & $-$2.79 & 0.53 & $-$16.64 & 1.19 & $-$8.38 & 0.35 & 0.78 & 0.45 & 0.64 & 8.92 \\
UBC 1 & 12.94 & 1.17 & 303.85 & 1.54 & 109.36 & 0.86 & $-$5.04 & 0.04 & $-$23.84 & 0.68 & $-$2.90 & 0.24 & 0.98 & 0.63 & 0.76 & 4.85 \\
ASCC 101 & 143.84 & 2.87 & 356.37 & 2.45 & 79.42 & 1.50 & $-$9.41 & 0.26 & $-$15.63 & 0.66 & $-$4.28 & 0.18 & 1.25 & 0.49 & 1.15 & 3.02 \\
Alessi 10 & 344.55 & 2.74 & 212.46 & 1.73 & $-$155.57 & 1.69 & $-$9.39 & 1.04 & $-$21.58 & 0.66 & $-$2.95 & 0.48 & 2.00 & 0.56 & 1.92 & 6.33 \\
UPK 88 & 119.26 & 3.09 & 251.96 & 2.56 & $-$60.43 & 2.66 & $-$4.38 & 0.48 & $-$3.93 & 1.08 & $-$14.73 & 0.40 & 3.35 & 3.34 & 0.29 & 2.39 \\
UBC 159 & $-$52.73 & 0.96 & 467.42 & 2.23 & $-$136.24 & 1.72 & 11.65 & 0.36 & $-$16.83 & 3.24 & $-$2.14 & 0.95 & 3.71 & 0.95 & 3.59 & 22.77 \\
Aveni$-$Hunter 1 & $-$141.72 & 2.36 & 380.74 & 4.25 & $-$91.55 & 4.22 & 6.30 & 1.66 & $-$15.33 & 4.73 & $-$0.34 & 1.35 & 1.09 & 0.76 & 0.79 & 10.80 \\}
\begin{landscape}
\begin{longtable}{l rr rr rr rr rr rr rrrr}
\caption{Cartesian coordinates for spatial and velocity coordinates together with their uncertainties.}\\
\hline
Cluster & $b_\mathrm{x}$ & $e_{b_\mathrm{x}}$ & $b_\mathrm{y}$ & $e_{b_\mathrm{y}}$ & $b_\mathrm{z}$ & $e_{b_\mathrm{z}}$ & $v_\mathrm{x}$ & $e_{v_\mathrm{x}}$ & $v_\mathrm{y}$ & $e_{v_\mathrm{y}}$ & $v_\mathrm{z}$  & $e_{v_\mathrm{z}}$ & $\sigma_{v2D}$ & $\sigma_{Vra}$ & $\sigma_{Vdec}$ & $\sigma_{Vrad}$ \\
\hline
 & pc & pc & pc & pc & pc & pc & km\,s$^{-1}$ & km\,s$^{-1}$ & km\,s$^{-1}$ & km\,s$^{-1}$ & km\,s$^{-1}$ & km\,s$^{-1}$ & km\,s$^{-1}$ & km\,s$^{-1}$ & km\,s$^{-1}$ & km\,s$^{-1}$ \\
\hline
\endfirsthead
\caption*{\textbf{Table B.3.} continued.} \\
\hline
Cluster & $b_\mathrm{x}$ & $e_{b_\mathrm{x}}$ & $b_\mathrm{y}$ & $e_{b_\mathrm{y}}$ & $b_\mathrm{z}$ & $e_{b_\mathrm{z}}$ & $v_\mathrm{x}$ & $e_{v_\mathrm{x}}$ & $v_\mathrm{y}$ & $e_{v_\mathrm{y}}$ & $v_\mathrm{z}$  & $e_{v_\mathrm{z}}$ & $\sigma_{v2D}$ & $\sigma_{Vra}$ & $\sigma_{Vdec}$ & $\sigma_{Vrad}$ \\
\hline
 & pc & pc & pc & pc & pc & pc & km\,s$^{-1}$ & km\,s$^{-1}$ & km\,s$^{-1}$ & km\,s$^{-1}$ & km\,s$^{-1}$ & km\,s$^{-1}$ & km\,s$^{-1}$ & km\,s$^{-1}$ & km\,s$^{-1}$ & km\,s$^{-1}$ \\
\hline
\endhead
\hline
Blanco 1 & 42.67 & 0.21 & 11.66 & 0.45 & $-$230.13 & 0.73 & $-$18.53 & 0.19 & $-$6.61 & 0.06 & $-$8.67 & 0.98 & 0.83 & 0.54 & 0.63 & 10.59 \\
UPK 303 & $-$162.03 & 1.28 & 118.56 & 1.43 & $-$64.87 & 1.39 & $-$5.75 & 3.52 & $-$29.11 & 2.68 & $-$10.68 & 1.41 & 1.26 & 0.67 & 1.06 & 15.29 \\
UPK 305 & $-$310.81 & 7.36 & 215.09 & 5.73 & $-$131.28 & 4.21 & $-$2.27 & 1.31 & $-$11.10 & 0.93 & $-$7.88 & 0.59 & 4.10 & 0.89 & 4.00 & 4.16 \\
NGC 1039 & $-$381.45 & 1.16 & 280.51 & 1.23 & $-$132.35 & 1.10 & 4.84 & 0.69 & $-$12.44 & 0.52 & $-$8.84 & 0.24 & 1.14 & 0.70 & 0.90 & 9.40 \\
Alessi 13 & $-$30.88 & 0.60 & $-$48.64 & 0.56 & $-$85.93 & 0.52 & $-$12.61 & 0.49 & $-$21.82 & 0.73 & $-$4.82 & 1.32 & 1.10 & 0.79 & 0.77 & 8.93 \\
Melotte 20 & $-$144.90 & 0.36 & 93.00 & 0.37 & $-$19.39 & 0.28 & $-$12.71 & 0.91 & $-$24.39 & 0.58 & $-$6.58 & 0.13 & 3.66 & 1.20 & 3.46 & 14.78 \\
Melotte 22 & $-$120.32 & 0.19 & 28.82 & 0.14 & $-$54.08 & 0.13 & $-$6.28 & 0.43 & $-$28.36 & 0.11 & $-$13.93 & 0.20 & 1.75 & 0.76 & 1.57 & 10.70 \\
UBC 31 & $-$330.25 & 1.37 & 99.08 & 0.73 & $-$89.49 & 0.62 & $-$18.25 & 1.63 & $-$5.90 & 0.49 & $-$6.61 & 0.45 & 7.04 & 2.93 & 6.40 & 16.31 \\
Melotte 25 & $-$43.16 & 0.12 & 0.26 & 0.07 & $-$17.21 & 0.06 & $-$41.91 & 0.34 & $-$19.28 & 0.06 & $-$1.18 & 0.17 & 3.67 & 2.57 & 2.62 & 6.59 \\
Platais 3 & $-$129.23 & 1.03 & 112.41 & 1.40 & 48.43 & 0.86 & $-$17.95 & 0.69 & $-$4.75 & 0.57 & $-$6.03 & 0.22 & 0.70 & 0.59 & 0.38 & 3.65 \\
NGC 1662 & $-$372.20 & 2.30 & $-$50.69 & 1.24 & $-$144.99 & 0.93 & 14.01 & 0.79 & 0.12 & 0.12 & 1.16 & 0.30 & 0.66 & 0.47 & 0.46 & 5.25 \\
UPK 385 & $-$296.07 & 5.10 & $-$41.48 & 2.57 & $-$92.67 & 1.80 & $-$12.67 & 2.70 & $-$7.69 & 0.40 & $-$4.76 & 0.88 & 1.71 & 0.62 & 1.59 & 6.41 \\
Platais 4 & $-$308.19 & 26.23 & $-$4.76 & 4.72 & $-$57.23 & 5.53 &   &   &   &   &   &   & 6.50 & 2.07 & 6.16 & 10.49 \\
NGC 1901 & 53.34 & 2.51 & $-$342.00 & 3.29 & $-$229.34 & 2.01 & $-$24.77 & 0.07 & $-$4.03 & 0.58 & $-$1.72 & 0.39 & 0.47 & 0.37 & 0.28 & 2.77 \\
Mamajek 3 & $-$90.82 & 2.03 & $-$27.77 & 1.07 & $-$25.96 & 0.56 & $-$11.34 & 0.16 & $-$19.63 & 0.43 & $-$8.55 & 0.03 & 0.29 & 0.14 & 0.25 & 0.50 \\
UBC 8 & $-$415.74 & 1.94 & 194.94 & 1.81 & 108.69 & 1.07 & 2.63 & 0.75 & $-$5.84 & 0.35 & $-$12.16 & 0.21 & 1.35 & 0.96 & 0.95 & 6.19 \\
UPK 350 & $-$392.91 & 3.44 & 88.30 & 2.98 & 141.47 & 1.86 & $-$6.92 & 0.72 & $-$7.34 & 0.18 & $-$12.59 & 0.28 & 0.37 & 0.23 & 0.29 & 2.67 \\
UBC 7 & $-$97.13 & 1.43 & $-$248.09 & 1.17 & $-$63.62 & 0.73 & $-$17.16 & 0.88 & $-$6.81 & 2.20 & $-$10.86 & 0.58 & 0.53 & 0.37 & 0.38 & 11.90 \\
Alessi 3 & $-$56.78 & 2.11 & $-$259.49 & 1.99 & $-$72.14 & 1.28 & $-$18.72 & 0.11 & 5.23 & 0.36 & $-$5.47 & 0.12 & 0.37 & 0.26 & 0.27 & 1.32 \\
Collinder 135 & $-$104.67 & 1.45 & $-$268.72 & 1.21 & $-$56.56 & 0.78 & $-$17.66 & 0.59 & $-$6.69 & 1.42 & $-$11.59 & 0.36 & 0.86 & 0.65 & 0.56 & 9.56 \\
UPK 540 & 4.54 & 2.06 & $-$346.19 & 2.73 & $-$106.25 & 1.61 & $-$15.50 & 0.03 & $-$5.56 & 2.54 & $-$3.89 & 0.79 & 3.66 & 1.80 & 3.19 & 10.22 \\
ASCC 41 & $-$221.59 & 2.06 & $-$181.30 & 1.22 & 62.66 & 1.12 & 10.95 & 0.91 & 2.03 & 0.75 & $-$3.91 & 0.26 & 0.44 & 0.30 & 0.31 & 6.16 \\
NGC 2516 & 26.20 & 0.32 & $-$390.93 & 0.47 & $-$111.19 & 0.27 & $-$21.81 & 0.04 & $-$23.93 & 0.25 & $-$4.10 & 0.08 & 2.44 & 1.43 & 1.97 & 6.46 \\
UPK 533 & $-$14.39 & 1.95 & $-$331.18 & 2.68 & $-$58.20 & 1.60 & $-$10.54 & 0.19 & $-$22.54 & 2.93 & $-$11.36 & 0.42 & 5.27 & 3.45 & 3.99 & 9.76 \\
UBC 12 & $-$260.05 & 5.96 & $-$330.07 & 6.49 & 122.07 & 3.44 & $-$33.37 & 0.58 & $-$15.59 & 0.77 & $-$1.15 & 0.25 & 2.94 & 2.52 & 1.50 & 3.50 \\
NGC 2632 & $-$139.27 & 0.28 & $-$67.66 & 0.21 & 98.20 & 0.24 & $-$42.69 & 0.20 & $-$20.39 & 0.10 & $-$9.28 & 0.14 & 1.64 & 1.20 & 1.12 & 5.17 \\
IC 2391 & 1.04 & 0.39 & $-$149.49 & 0.49 & $-$17.77 & 0.35 & $-$23.71 & 0.06 & $-$13.25 & 1.17 & $-$5.46 & 0.15 & 1.75 & 0.84 & 1.54 & 11.17 \\
Mamajek 1 & 35.04 & 0.79 & $-$87.26 & 1.41 & $-$36.61 & 0.78 & $-$12.13 & 0.27 & $-$18.31 & 1.07 & $-$10.46 & 0.39 & 3.47 & 2.18 & 2.69 & 4.15 \\
Turner 5 & $-$41.94 & 2.51 & $-$406.95 & 3.11 & 79.37 & 3.07 & 4.23 & 0.14 & 2.03 & 0.58 & $-$4.78 & 0.22 & 1.27 & 0.37 & 1.22 & 2.35 \\
IC 2602 & 50.27 & 0.33 & $-$141.16 & 0.62 & $-$12.77 & 0.35 & $-$8.02 & 0.50 & $-$21.56 & 1.39 & $-$0.81 & 0.14 & 2.11 & 0.65 & 2.01 & 15.59 \\
Melotte 111 & $-$7.13 & 0.15 & $-$6.20 & 0.35 & 85.01 & 0.52 & $-$2.41 & 0.06 & $-$5.50 & 0.06 & $-$1.28 & 0.65 & 0.43 & 0.32 & 0.29 & 4.50 \\
UPK 606 & 123.36 & 1.16 & $-$107.42 & 1.46 & 39.55 & 1.35 & $-$7.11 & 3.62 & $-$20.25 & 3.15 & $-$5.08 & 1.17 & 9.95 & 9.42 & 3.20 & 17.12 \\
UPK 612 & 172.40 & 0.97 & $-$131.87 & 0.98 & $-$59.83 & 0.56 & $-$9.14 & 0.62 & $-$17.09 & 0.45 & 3.25 & 0.23 & 2.86 & 1.75 & 2.27 & 5.91 \\
UPK 624 & 262.67 & 2.86 & $-$157.49 & 2.61 & $-$58.48 & 1.51 & $-$9.08 & 1.71 & $-$15.84 & 1.04 & $-$11.34 & 0.41 & 5.33 & 1.25 & 5.18 & 6.77 \\
Alessi 24 & 396.28 & 3.13 & $-$238.22 & 3.36 & $-$120.50 & 2.01 & $-$4.77 & 0.62 & $-$17.94 & 0.39 & $-$11.21 & 0.22 & 0.65 & 0.43 & 0.49 & 3.24 \\
Mamajek 2 & 182.07 & 2.04 & 55.87 & 1.05 & 42.19 & 0.52 & $-$10.64 & 1.85 & $-$26.10 & 0.57 & $-$4.88 & 0.46 & 1.14 & 0.68 & 0.91 & 7.51 \\
IC 4665 & 280.57 & 2.73 & 165.23 & 1.42 & 100.92 & 0.67 & $-$3.09 & 1.47 & $-$17.15 & 0.88 & $-$8.52 & 0.53 & 0.62 & 0.46 & 0.41 & 8.34 \\
Collinder 350 & 315.85 & 2.86 & 159.85 & 1.46 & 92.99 & 0.67 & $-$13.04 & 1.25 & $-$11.09 & 0.64 & 3.23 & 0.38 & 0.52 & 0.42 & 0.30 & 7.10 \\
Gulliver 20 & 318.49 & 4.05 & 256.38 & 2.77 & 95.29 & 1.11 & $-$4.15 & 1.18 & $-$17.23 & 0.93 & $-$10.93 & 0.33 & 3.06 & 1.14 & 2.84 & 6.86 \\
Mamajek 4 & 405.21 & 2.37 & $-$117.44 & 2.06 & $-$129.03 & 1.25 & $-$45.18 & 0.59 & $-$25.62 & 0.23 & $-$15.82 & 0.23 & 1.94 & 1.31 & 1.44 & 3.93 \\
UBC 9 & 190.79 & 2.69 & 268.12 & 1.82 & 99.16 & 0.96 & $-$3.92 & 1.08 & $-$18.69 & 1.50 & $-$9.79 & 0.58 & 1.89 & 0.42 & 1.85 & 8.64 \\
UPK 654 & 487.26 & 6.07 & $-$4.07 & 2.12 & $-$92.59 & 1.56 & $-$0.19 & 4.11 & $-$17.25 & 0.11 & $-$11.60 & 0.82 & 4.14 & 1.16 & 3.98 & 16.71 \\
Stephenson 1 & 134.26 & 1.60 & 314.47 & 1.31 & 93.74 & 0.77 & $-$2.79 & 0.53 & $-$16.64 & 1.19 & $-$8.38 & 0.35 & 0.78 & 0.45 & 0.64 & 8.92 \\
UBC 1 & 12.94 & 1.17 & 303.85 & 1.54 & 109.36 & 0.86 & $-$5.04 & 0.04 & $-$23.84 & 0.68 & $-$2.90 & 0.24 & 0.98 & 0.63 & 0.76 & 4.85 \\
ASCC 101 & 143.84 & 2.87 & 356.37 & 2.45 & 79.42 & 1.50 & $-$9.41 & 0.26 & $-$15.63 & 0.66 & $-$4.28 & 0.18 & 1.25 & 0.49 & 1.15 & 3.02 \\
Alessi 10 & 344.55 & 2.74 & 212.46 & 1.73 & $-$155.57 & 1.69 & $-$9.39 & 1.04 & $-$21.58 & 0.66 & $-$2.95 & 0.48 & 2.00 & 0.56 & 1.92 & 6.33 \\
UPK 88 & 119.26 & 3.09 & 251.96 & 2.56 & $-$60.43 & 2.66 & $-$4.38 & 0.48 & $-$3.93 & 1.08 & $-$14.73 & 0.40 & 3.35 & 3.34 & 0.29 & 2.39 \\
UBC 159 & $-$52.73 & 0.96 & 467.42 & 2.23 & $-$136.24 & 1.72 & 11.65 & 0.36 & $-$16.83 & 3.24 & $-$2.14 & 0.95 & 3.71 & 0.95 & 3.59 & 22.77 \\
Aveni$-$Hunter 1 & $-$141.72 & 2.36 & 380.74 & 4.25 & $-$91.55 & 4.22 & 6.30 & 1.66 & $-$15.33 & 4.73 & $-$0.34 & 1.35 & 1.09 & 0.76 & 0.79 & 10.80 \\
\hline
\label{tab.clusters_characteristics_cartesian}
\end{longtable}
\tablefoot{Columns $\sigma_{v2D}$, $\sigma_{Vra}$, $\sigma_{Vdec}$, and $\sigma_{Vrad}$ list velocity dispersions for 2D, transversal, and line-of-sight velocities.}
\end{landscape}
\twocolumn

\FloatBarrier
\section{Catalogue of cluster members}
We describe a schema of the table with cluster members in \ref{tab.catalog_members_format}. We note that some \texttt{source\_id} entries are repeated. This happens in cases where stars are simultaneously candidate members of two different clusters that overlap in their outskirts, for example, in the case of the known pair Collinder\,135 and UBC\,7 (see Appendices \ref{sec.pair} and \ref{sec.triplet}).

\onecolumn
\begin{longtable}{llll}
\caption{Table schema for the catalogue of cluster members.}\\
\hline
Column Name & Units & Description & Data Type \\
\hline
\endfirsthead
\caption{continued.}\\
\hline
Column Name & Units & Description & Data Type \\
\hline
\endhead
\hline
\multicolumn{4}{c}
{\tablefoot{The entire table for all clusters in this work is available only in electronic form at the CDS via anonymous ftp to cdsarc.cds.unistra.fr (130.79.128.5) or via https://cdsarc.cds.unistra.fr/cgi-bin/qcat?J/A+A/ as well as at our website described in Section\,\ref{sec.website}.}} \\ 
\endfoot
source\_id &  & \textit{Gaia}\,DR3 identifier & integer \\
ra & deg & propagated from \textit{Gaia}\,DR3 (or Hipparcos if HIP$>$0) & float \\
dec & deg & propagated from \textit{Gaia}\,DR3 (or Hipparcos if HIP$>$0) & float \\
pmra & mas\,yr$^{-1}$ & propagated from \textit{Gaia}\,DR3 (or Hipparcos if HIP$>$0) & float \\
pmra\_error & mas\,yr$^{-1}$ & propagated from \textit{Gaia}\,DR3 (or Hipparcos if HIP$>$0) & float \\
pmdec & mas\,yr$^{-1}$ & propagated from \textit{Gaia}\,DR3 (or Hipparcos if HIP$>$0) & float \\
pmdec\_error & mas\,yr$^{-1}$ & propagated from \textit{Gaia}\,DR3 (or Hipparcos if HIP$>$0) & float \\
parallax & mas & propagated from \textit{Gaia}\,DR3 (or Hipparcos if HIP$>$0) & float \\
parallax\_error & mas & propagated from \textit{Gaia}\,DR3 (or Hipparcos if HIP$>$0) & float \\
r\_med\_geo & pc & propagated from 2021AJ....161..147B & float \\
r\_lo\_geo & pc & propagated from 2021AJ....161..147B & float \\
r\_hi\_geo & pc & propagated from 2021AJ....161..147B & float \\
radial\_velocity & km\,s$^{-1}$ & propagated from \textit{Gaia}\,DR3 & float \\
radial\_velocity\_error & km\,s$^{-1}$ & propagated from \textit{Gaia}\,DR3 & float \\
\hline
HIP &  & Hipparcos identifier & integer \\
ruwe &  & propagated from \textit{Gaia}\,DR3 & float \\
WD &  & Star in white dwarf catalogue from 2021MNRAS.508.3877G & boolean \\
binary1 &  & binary star (primary) in 2021MNRAS.506.2269E & boolean \\ 
binary2 &  & binary star (secondary) in 2021MNRAS.506.2269E & boolean \\
\hline
phot\_g\_mean\_mag & mag & propagated from \textit{Gaia}\,DR3 & float \\
phot\_bp\_mean\_mag & mag & propagated from \textit{Gaia}\,DR3 & float \\
phot\_rp\_mean\_mag & mag & propagated from \textit{Gaia}\,DR3 & float \\
Hpmag\_HIP & mag & Hipparcos magnitude & float \\
B-V & mag & propagated from Hipparcos & float \\
G-Rp & mag & propagated from \textit{Gaia}\,DR3 & float \\
designation\_2mass & & propagated from 2006AJ....131.1163S &  \\
j\_m & mag & propagated from 2006AJ....131.1163S & float \\
j\_msigcom & mag & propagated from 2006AJ....131.1163S & float \\
h\_m & mag & propagated from 2006AJ....131.1163S & float \\
h\_msigcom & mag & propagated from 2006AJ....131.1163S & float \\
ks\_m & mag &propagated from 2006AJ....131.1163S  & float \\
ks\_msigcom & mag & propagated from 2006AJ....131.1163S & float \\
p\_obj\_id &  & propagated from 2020ApJS..251....7F &  \\
g\_mean\_psf\_mag & mag & propagated from 2020ApJS..251....7F & float \\
g\_mean\_psf\_mag\_error & mag & propagated from 2020ApJS..251....7F & float \\
r\_mean\_psf\_mag & mag & propagated from 2020ApJS..251....7F & float \\
r\_mean\_psf\_mag\_error & mag & propagated from 2020ApJS..251....7F & float \\
i\_mean\_psf\_mag & mag & propagated from 2020ApJS..251....7F & float \\
i\_mean\_psf\_mag\_error & mag & propagated from 2020ApJS..251....7F & float \\
z\_mean\_psf\_mag & mag & propagated from 2020ApJS..251....7F & float \\
z\_mean\_psf\_mag\_error & mag & propagated from 2020ApJS..251....7F & float \\
y\_mean\_psf\_mag & mag & propagated from 2020ApJS..251....7F & float \\
y\_mean\_psf\_mag\_error & mag & propagated from 2020ApJS..251....7F & float \\
allwise\_designation &  & propagated from 2010AJ....140.1868W &  \\
w1mpro & mag & propagated from 2010AJ....140.1868W & float \\
w1mpro\_error & mag & propagated from 2010AJ....140.1868W & float \\
w2mpro & mag & propagated from 2010AJ....140.1868W & float \\
w2mpro\_error & mag & propagated from 2010AJ....140.1868W & float \\
w3mpro & mag & propagated from 2010AJ....140.1868W & float \\
w3mpro\_error & mag & propagated from 2010AJ....140.1868W & float \\
w4mpro & mag & propagated from 2010AJ....140.1868W & float \\
w4mpro\_error & mag & propagated from 2010AJ....140.1868W & float \\
sdss\_objid &  & propagated from 2017ApJS..233...25A &  \\
u & mag & propagated from 2017ApJS..233...25A & float \\
err\_u & mag & propagated from 2017ApJS..233...25A & float \\
g & mag & propagated from 2017ApJS..233...25A & float \\
err\_g & mag & propagated from 2017ApJS..233...25A & float \\
r & mag & propagated from 2017ApJS..233...25A & float \\
err\_r & mag & propagated from 2017ApJS..233...25A & float \\
i & mag & propagated from 2017ApJS..233...25A & float \\
err\_i & mag & propagated from 2017ApJS..233...25A & float \\
z & mag & propagated from 2017ApJS..233...25A & float \\
err\_z & mag & propagated from 2017ApJS..233...25A & float \\
JName\_ukidss & mag & propagated from 2007MNRAS.379.1599L &  \\            
zAperMag3 & mag & propagated from 2007MNRAS.379.1599L & float \\
zAperMag3Err & mag & propagated from 2007MNRAS.379.1599L & float \\
yAperMag3 & mag & propagated from 2007MNRAS.379.1599L & float \\
yAperMag3Err & mag & propagated from 2007MNRAS.379.1599L & float \\
jAperMag3 & mag & propagated from 2007MNRAS.379.1599L & float \\
jAperMag3Err & mag & propagated from 2007MNRAS.379.1599L & float \\
hAperMag3 & mag & propagated from 2007MNRAS.379.1599L & float \\
hAperMag3Err & mag & propagated from 2007MNRAS.379.1599L & float \\
k\_1AperMag3 & mag & propagated from 2007MNRAS.379.1599L & float \\
k\_1AperMag3Err & mag & propagated from 2007MNRAS.379.1599L & float \\
SrcID\_vista & mag & propagated from 2012sngi.confE..37M & float \\
Ypmag & mag & propagated from 2012sngi.confE..37M & float \\
e\_Ypmag & mag & propagated from 2012sngi.confE..37M & float \\
Jpmag & mag & propagated from 2012sngi.confE..37M & float \\
e\_Jpmag & mag & propagated from 2012sngi.confE..37M & float \\
Hpmag & mag & propagated from 2012sngi.confE..37M & float \\
e\_Hpmag & mag & propagated from 2012sngi.confE..37M & float \\
Kspmag & mag & propagated from 2012sngi.confE..37M & float \\
e\_Kspmag & mag & propagated from 2012sngi.confE..37M & float \\
\hline
Cluster &  & Parent cluster & string \\
mass & M$_\odot$ & Mass estimate from this work & float \\
bx & pc & Heliocentric cartesian coordinate X & float \\
by & pc & Heliocentric cartesian coordinate Y & float \\
bz & pc & Heliocentric cartesian coordinate Z & float \\
vx & km\,s$^{-1}$ & Heliocentric cartesian velocity in X direction & float \\
vy & km\,s$^{-1}$ & Heliocentric cartesian velocity in Y direction & float \\
vz & km\,s$^{-1}$ & Heliocentric cartesian velocity in Z direction & float \\
c &  & A proxy for membership probability & float \\
dist2center & pc & Distance from the cluster centre & float\\
bona\_fide &  & Bona fide member & boolean \\
Ntr &  & Distance from the cluster centre in tidal radii & integer \\
\label{tab.catalog_members_format}
\end{longtable}
\twocolumn

\FloatBarrier
\section{Individual clusters} \label{sec.individual_clusters}

\subsection{A pair: Collinder 135 and UBC 7} \label{sec.pair}
Clusters Collinder\,135 and UBC\,7 are a physical pair of clusters \citep{2020A&A...642L...4K}. They have very similar proper motions, but their distances to the Sun are 294\,pc and 273\,pc, respectively. 
Due to their proximity -- they are found only 23\,pc apart -- some of their candidate members at 3\,tidal radii overlap (52 stars).

\subsection{A triplet: NGC 2516, UPK 540, and UPK 533} \label{sec.triplet}
The Southern Beehive (NGC\,2516) is known to be enveloped by a large 500\,pc halo
\citep{2021AJ....162..197B}. While we only list candidates up to its 3\,tidal radii, we find a partial overlap between Southern Beehive, UPK\,540, and UPK\,533. While they lie 54\,pc apart from each other, UPK\,540 and UPK\,533 are found 50 and 90\,pc away from NGC\,2516, respectively.

Out of 54 bona fide members in UPK\,540, 25 are found in the halo of NGC\,2516 in \citep{2021AJ....162..197B}, while there are no stars from UPK\,533 in the halo.
At the same time, UPK\,533 is located very close to another cluster, NGC\,2547. They are close but clearly separable in the proper motion space, but we found a significant overdensity of stars from NGC\,2547 within the 3\,tidal radii of UPK\,533. Based on the colour-magnitude diagram of UPK\,533, the contamination of its 44 bona fide members is likely small.

Due to their small separation, our list of members includes repeated stars that belong to two clusters. There are 158 repeated stars from the pair UPK\,540--NGC\,2516 and five from the pair UPK\,540--UPK\,533.

\subsection{Platais 4} \label{sec.platais4}
This cluster is a weak overdensity next to a massive cluster in the proper motion space. Both clusters are located in a similar region of the sky. We believe that these factors led to a false characterisation of this cluster. The results for this cluster are not convincing, especially because we do not have a 3D velocity for it.

\subsection{UBC 31}
This cluster is located in a very complex region of the proper motion space with no clear distinction from its close neighbours, NGC\,1579 and IC\,348. The bona fide members of this cluster suffer from a higher contamination rate that increases significantly for candidate members beyond 1 tidal radius.

\subsection{UBC 159}
One of the most distant clusters in our list, UBC\,159 is located at around 500\,pc from the Sun. This is a young cluster with its members in the pre-main-sequence stage. Due to its large distance, the contamination rate seems to be higher than for other clusters.


\end{appendix}
\end{document}